\documentclass[pra,preprint,11pt,superscriptaddress]{revtex4-1} % On govt laptop
\usepackage{amsmath}
\usepackage{amssymb}
\usepackage{graphicx}
\usepackage{color}
\newcommand{\be}[1]{\begin{equation}\label{#1}}
\newcommand{\ee}{\end{equation}}
\newcommand{\bea}[1]{\begin{eqnarray}\label{#1}}
\newcommand{\eea}{\end{eqnarray}}
\newcommand{\no}{\nonumber \\}
\newcommand{\Fig}[1]{Fig.(\ref{#1})}
\newcommand{\Eq}[1]{Eq.(\ref{#1})}
\newcommand{\App}[1]{Appendix~\ref{#1}}
\newcommand{\Sec}[1]{Section~\ref{#1}}
\newcommand{\bsub}{\begin{subequations}}
\newcommand{\esub}{\end{subequations}}
\newcommand{\Int}[1]{\int_{0}^{#1 L}dz\,e^{i\xi(\omega)[#1 L - z]} \hat{s}(z,\omega)}
\newcommand{\om}{\omega}

\def\a0{{\alpha_0}}

\def\da0{{\dot{\alpha}_0}}

\def\myoverDefn#1#2{\hbox{\space \raise-2mm\hbox{$\textstyle{#1} \atop \scriptstyle{#2}$} }}
\def\ha{{\hat{a}}}
\def\haint{{\hat{a}_{int}}}
\def\hc{{\hat{c}}}
\def\hf{{\hat{f}}}
\def\hb{{\hat{b}}}
\def\hd{{\hat{d}}}
\def\hF{{\hat{F}}}
\def\hs{{\hat{s}}}
\def\om{{\omega}}
\def\k{{\kappa}}
\def\t{{\tau}}
\def\ts{{\tau^{*}}}
\def\ks{{\kappa^{*}}}
\def\mA{{\mathcal{A}}}
\def\mB{{\mathcal{B}}}
\def\gc{{\gamma_{c}}}
\def\gint{{\gamma_{int}}}
\def\gp{{\gamma_{+}}}
\def\gm{{\gamma_{-}}}
\def\G{{\Gamma}}
\def\g{{\gamma}}
\def\d{{\delta}}
\def\a{{\alpha}}
\def\M{\mathcal{M}}
\newcommand{\ket}[1]{|#1\rangle}
\newcommand{\bra}[1]{\langle #1|}

\begin{document}
%============================================
%\flushleft{\large DRAFT: rr\_losses\_paper\_v3.tex}
%\vspace{-1em}\flushleft{(\textit{v1: original rough draft; v2: references added; v3: (somewhat) reduced appendices})}
%\vspace{-1em}\flushleft{({\textit{\color{red} Conclusion still needed})}
%% _v4.tex PMA version to all atuhors 18July2016
%% _v5.tex AMS edits 21July2016
%% _v6.tex PMA edits and final proof read - still need to check figure captions
%============================================
\title{A quantum optical description of losses in ring resonators \\ based on field operator transformations}
\author{Paul M. Alsing}
\affiliation{Air Force Research Laboratory, Information Directorate, 525 Brooks Rd, Rome, NY, 13411}
\author{Edwin E. Hach III}
\affiliation{Rochester Institute of Technology, School of Physics and Astronomy, 85 Lomb Memorial Dr. Rochester, NY 14623}
\author{Christopher C. Tison}
\affiliation{Air Force Research Laboratory, Information Directorate, 525 Brooks Rd, Rome, NY, 13411}
\affiliation{Quanterion Solutions Incorporated, 100 Seymour Rd, Utica, NY 13502}
\author{A. Matthew Smith}
\affiliation{Air Force Research Laboratory, Information Directorate, 525 Brooks Rd, Rome, NY, 13411}
\date{\today}
%\maketitle

\begin{abstract}
In this work we examine loss in ring resonator networks from an ``operator valued phasor addition" approach (or OVPA approach) which considers the multiple transmission and cross coupling paths of a quantum field traversing a ring resonator coupled to one or two external waveguide buses. We demonstrate the consistency of our approach by the preservation of the operator commutation relation of the out-coupled bus mode.
We compare our results to those obtained from the conventional quantum Langevin approach which introduces
noise operators in addition to the quantum Heisenberg equations in order to preserve commutation
relations in the presence of loss.
It is shown that the two expressions agree in the neighborhood of a cavity resonance where the Langevin approach is applicable, whereas the operator valued phasor addition expression we derive is more general, remaining valid far from resonances.
 In addition, we examine the effects of internal and coupling losses on the Hong-Ou-Mandel manifold first discussed in Hach \textit{et al.} Phys. Rev. A \textbf{89}, 043805 (2014) that generalizes the destructive interference of two incident photons interfering on a 50:50 beam splitter (HOM effect) to the case of an add/drop double bus ring resonator.
\end{abstract}
\maketitle
%==================================
\section{Introduction}
%==================================
It is difficult to overstate the importance of the control of fields at the single or few photon level in the realization of optical architectures for quantum computation, communication, and metrology. In order to optimize the functionality of next-generation quantum information processing systems, devices need to be scaled to the level of micro- or even nano-integration. Notable persistent challenges to advancement of efficient, scalable quantum information processing systems include the identification of useful physical qubits, the discovery of materials for use in quantum circuits, and the development of system architectures based on those qubits and materials. Light-speed transmission and high resilience to noise in comparison with other possible physical systems identifies photons as a very promising realization of the carriers of quantum (and classical) information. Further, several degrees of freedom, for example, presence/absence of a photon or mutually orthogonal optical polarization states can be used to encode quantum information \cite{Kok_Lovett:2010}.

One potential platform is silicon, which has desirable optical properties for integrated optical systems at the telecommunication wavelength of 1550 nm. In addition, silicon is a candidate for fabricating sub-Poissonian single photon sources relying on its high third order nonlinearity $\chi^3$ \cite{Clemmen:2009}. Using such sources, several diverse and exciting quantum phenomena can be explored, including time bin entanglement \cite{Marcikic:2004}, polarization entanglement \cite{Li:2005}, and N00N reduced de-Broglie wavelength \cite{Preble:2015}. Pioneered largely by the early work of Yariv \cite{Yariv:2000}, silicon micro-ring resonators evanescently coupled to silicon wave guides \cite{Bogaerts:2012} find an ever-growing range of applications as the bases for devices and networks that are at the heart of the phenomena underpinning many quantum technologies
%[menu of: Sipe, Stefan, et al, Our 2014 paper, Obrien, Englund, Lipson, Harris, etc?].
\cite{Preble:2015,Sipe:2015a,Sipe:2015b,OBrien:2015,Hach:2014,Englund:2014,Lipson:2008}.
In particular, our collaboration has recently demonstrated theoretically a particular
enhancement of the Hong-Ou-Mandel Effect \cite{Hach:2014} and experimentally a two-photon interference effect in down converted photons generated on-chip in a silicon microring resonator \cite{Preble:2015,OBrien:2015}.

Naturally paralleling the increased interest in silicon microring resonator networks, a significant body of theoretical analysis has developed into a reasonably sophisticated description of the quantum optical transport behaviors exhibitied in various simple topologies and environments. Two basic approaches have emerged in formulating the theoretical description of such systems. One that we shall refer to as the Langevin approach is based upon Lipmann-Schwinger style scattering theory at the localized couplers between components (i.e. microrings and waveguides) along with photonic losses modeled via noise operators representing a thermal bath of oscillators
\cite{Shen_Fan:2007,*Shen_Fan:2009a,*Shen_Fan:2009b,Tsang:2010,*Tsang:2011,Hach:2010,Agarwal:2014,Sipe:2015b,Shapiro:2015}.
%
%[Shen and Fan, Our 2010 Paper, Sipe no free lunch, tsang, other langevin stuff].
%\cite{Shen_Fan:2007,*Shen_Fan:2009a,*Shen_Fan:2009b} \cite{Tsang:2010,*Tsang:2011} \cite{Hach:2010,Sipe:2015b,Shapiro:2015}.
%
The second approach, which we describe below, which we will loosely call ``operator valued phasor addition" or the OVPA approach, is based upon the construction of field transformations for the optical mode operators by considering a linear superposition of transition amplitudes through all possible paths of the optical system %\cite{Skaar:2004,Hach:2013,Hach:2014}  \cite{Ataman:2014a,*Ataman:2014b,*Ataman:2015a,*Ataman:2015b} \cite{Hach:2015,*Hach:2016}.
\cite{Loudon:1995,*Loudon:1996,Skaar:2004,Hach:2013,Hach:2014,Ataman:2014a,*Ataman:2014b,*Ataman:2015a,*Ataman:2015b,Hach:2015,*Hach:2016}

The Langevin approach \cite{Walls_Milburn:1994,Mandel_Wolf:1995,Scully_Zubairy:1997,Orszag:2000} is advantageous with respect to its natural incorporation of quantum noise and its seamless incorporation of finite coherence times and bandwidths. The significant disadvantages of the Langevin approach are that it is
%prohibitively
difficult to apply to photonic input states that are more exotic than one or two-photon Fock states and that it oversimplifies to some degree the topology of the ring, potentially creating stumbling blocks in the analysis of larger quantum networks of microrings and waveguides. %The second approach, which we loosely call the ``operator valued phasor" approach is based upon the construction of the input-output propagator for the optical mode operators by considering a linear superposition of transition amplitudes through all possible paths of the optical system
%[Us 2014, skaar, ataman, my SPIE papers].
%%\cite{Ataman:2014a,*Ataman:2014b,*Ataman:2015a,*Ataman:2015b}
%%\cite{Skaar:2004,Hach:2013,Hach:2014,*Ataman:2014a,*Ataman:2014b,*Ataman:2015a,*Ataman:2015b,Hach:2015,Hach:2016}.
%\cite{Skaar:2004,Hach:2013,Hach:2014}  \cite{Ataman:2014a,*Ataman:2014b,*Ataman:2015a,*Ataman:2015b} \cite{Hach:2015,*Hach:2016}.
Our OVPA approach is based on input and output states of the quantum optical system which are related by working in an effective Heisenberg picture \cite{Gerry_Knight:2004}. This approach is easy to generalize to all network topologies and arbitrary photonic input states. Previous works along this line of analysis have focused almost entirely upon lossless %continuous wave (cw)
operation of the networks \cite{Loudon:1995,*Loudon:1996,Hach:2014,Ataman:2014a,*Ataman:2014b,*Ataman:2015a,*Ataman:2015b}.  These previous works have yielded interesting results, even within the confines of such idealized conditions.
The principal result of this present work is to extend the analysis of silicon microring resonator networks to larger and more general devices.  We formulate an approach capable of capturing the
advantages of both of the Langevin and previous operator multi-path approaches in this area. %This unification of approaches is precisely what we accomplish in this paper.

The paper is organized as follows.
% Section II
In \Sec{derivation_single_bus} we derive the internal cavity and output mode of an all through ring resonator (often called a single bus ring resonator) from the conventional quantum Langevin approach which entails the inclusion of quantum noise bath operators. We relate the expression for the out-coupled mode (exiting the bus) to the expression found by considering the phasor addition of multiple transmission and cross coupling paths of a classical field traversing the ring resonator. This latter classical approach is equivalent to considering the junction of the bus to the ring resonator as an effective transmission/reflection beam splitter interaction with cross coupling acting
analogously as an effective ``reflection"  of the external bus driving field into the ring resonator.
% Section III

In \Sec{sec:pma:rr:deriv} we quantize the OVPA approach. Unlike other multi-path approaches considered in literature, we explicitly include quantum noise using Loudon's expression for attenuation loss of a traveling wave mode \cite{Loudon:1997,Loudon:2000}, now adapted to the ring resonator/bus geometries. The expression for the single bus resonator output mode is  compared to the corresponding expression derived from the Langevin approach in \Sec{derivation_single_bus}. It is shown that the two expressions agree in the neighborhood of a cavity resonance where the Langevin approach is applicable. The OVPA expression we derive is more general, remaining valid  far from resonance. We also generalize our OVPA approach to the case of the add/drop (or double bus) ring resonator.

% Section IV
In \Sec{sec:HOMM_with_loss} we examine the effects of internal and coupling losses on the Hong-Ou-Mandel manifold first discussed in Hach \textit{et al.} \cite{Hach:2014} that generalizes the destructive interference of two incident photons interfering on a 50:50 beam splitter (HOM effect \cite{HOM:1987}) to the case of an add/drop double bus ring resonator.
% Section V
In \Sec{conclusion} we state our conclusions and outlook for future work.

To make this paper self contained we relegate many of the algebraic and background details to the appendices.
% Appendix A
In \App{app:haus} we review the classical derivation of the input-output formalism that is used in this work.
% to relate the internal cavity mode of the ring resonator to the input driving and out-coupled bus modes.
%
% Appendix B
In \App{app:walls_milburn} we review the quantum derivation of the input-output formalism, where the emphasis is on the preservation of the operator commutation relations. % required by the unitarity of the quantum evolution.
%
% Appendix C
In \App{app:loudon} we review Loudon's quantum formulation of traveling-wave attenuation in a beam that we adapt in the main body of the text to the ring resonator geometries. % considered, and explicitly demonstrate the preservation of the commutation relations for both uniform and piecewise uniform loss within the ring resonator.
%
% Appendix D
In \App{app:pma:comm:deriv} we explicitly demonstrate the quantum commutation relation for the expression for the out-coupled single bus mode.

%==================================
\section{Derivation of output field of an all through (single bus) ring resonator}\label{derivation_single_bus}
%==================================
\subsection{Langevin approach derivation}
\begin{figure}[h]
%\begin{tabular}{cc}
\includegraphics[width=3.0in,height=2.5in]{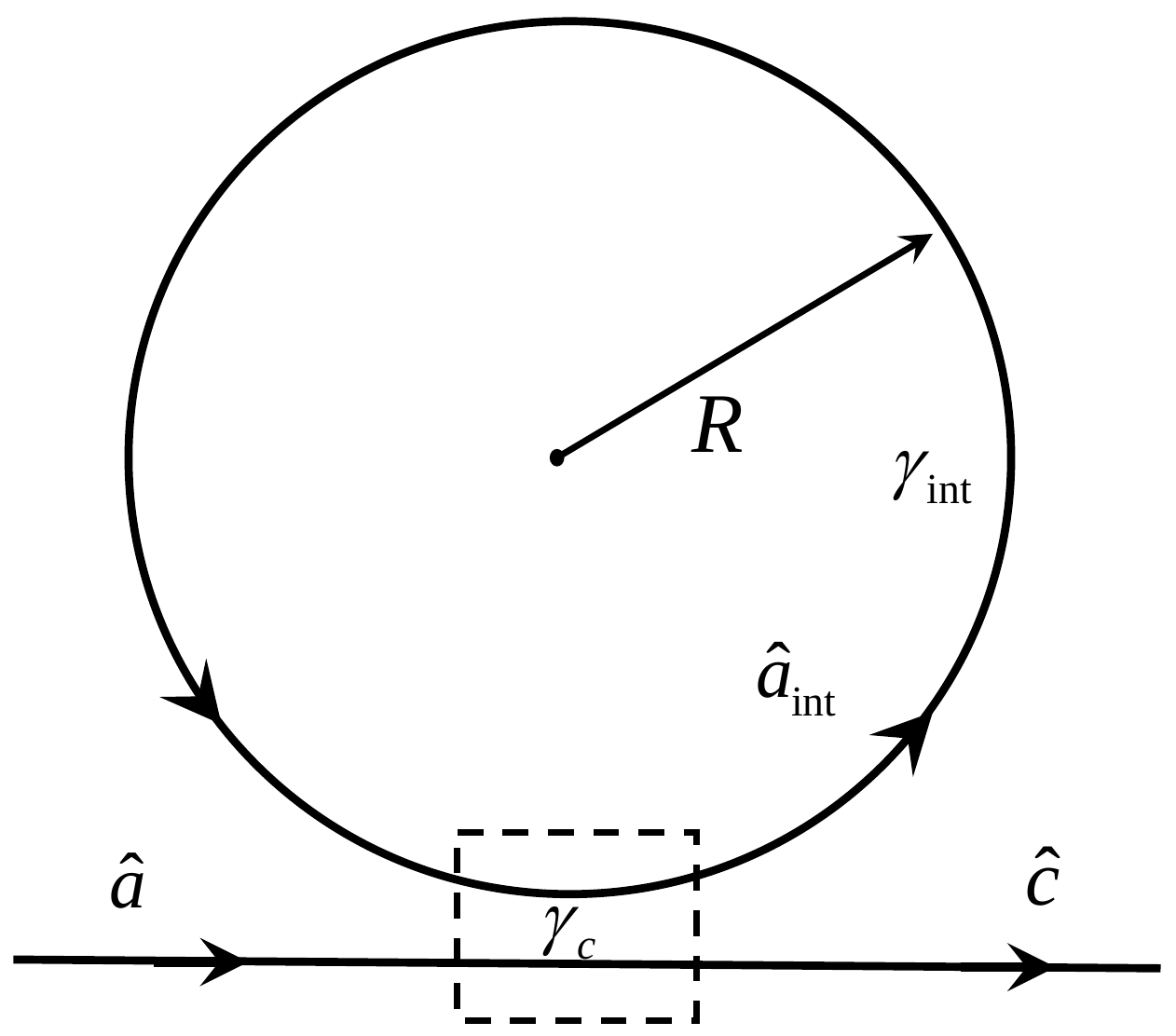}
%\includegraphics[width=3.0in,height=2.5in]{rr_losses_all_thru_RR_a_c_aint} %& Original Figure Name
%\includegraphics[width=3.0in,height=2.5in]{rr_loss_all_thru_RR_a_c}
%\end{tabular}
\caption{An all through (single bus) ring resonator
}\label{fig:opr:a:c:aint}
\end{figure}

%==================================
In this section we follow a conventional Langevin approach \cite{Walls_Milburn:1994,Mandel_Wolf:1995,Scully_Zubairy:1997,Orszag:2000}
for the derivation of the output field of an single bus ring resonator.
In \Fig{fig:opr:a:c:aint} we show a microring resonator with input (quantized) field $\hat{a}$, output field $\hat{c}$, and internal ring resonator cavity mode $\ha_{int}$. Here, $\gamma_c$ is the coupling coefficient between the input and internal mode and $\gamma_{int}$ represents internal losses.
%==================================
%==================================
Following the derivation in \Eq{eqn:aindot:ain:f} in \App{app:walls_milburn}
the equation of motion for the driven internal field $\ha_{int}$ undergoing coupling and internal losses is given by,
\be{eqn:opr:aintdot:a:f}
\dot{\hat{a}}_{int}(t) = -\frac{i}{\hbar}\,[\hat{a}_{int}, H_{sys}]
                       - \frac{(\gamma_c+\gamma_{int})}{2}\,\hat{a}_{int}(t)
                       + \sqrt{\gamma_c}\,\hat{a}(t)
                       + \sqrt{\gamma_{int}}\,\hat{f}(t),
\ee
where $\hat{f}(t)$ are the quantum Langevin noise operators satisfying the white
noise commutation relations $[\hat{f}(t), \hat{f}^\dagger(t')] = \delta(t-t')$.
As discussed in \App{app:walls_milburn} their presence is required by quantum mechanics to ensure that the commutation relations for the internal field are satisfied.
In addition to \Eq{eqn:opr:aintdot:a:f}, a boundary condition between the input, output and internal field is given by,
\be{eqn:Langevin:input:output:BC}
\hat{a} + \hat{c} = \sqrt{\gamma_c}\,\hat{a}_{int}.
\ee
This boundary condition follows from the widely used \textit{input-output} formalism \cite{Collett_Gardiner:1984,Walls_Milburn:1994,Mandel_Wolf:1995,Scully_Zubairy:1997,Orszag:2000}, a quantum optical instantiation of the $S$ matrix theory, relating early time input fields to late time output fields in scattering problems.
We present the derivation  of \Eq{eqn:Langevin:input:output:BC} classically in \App{app:haus}, and quantum mechanically in \App{app:walls_milburn}. In quantum optics, this boundary condition is used to related the internal cavity mode $\ha_{int}$ to the external driving $\ha$ and out-coupled $\hc$ modes.

For simplicity we take $H_{sys} = \hbar\omega_0 \ha^\dagger_{int} \ha_{int}$ to be the free field Hamiltonian for the internal ring resonator mode of frequency $\omega_0$. Transforming to the frequency domain via
$\hat{a}_{int}(t) = \int_{-\infty}^{\infty}d\omega \,\hat{a}_{int}(\omega)\,e^{-i\omega t}$
yields,
\be{soln:aint:Langevin}
\hat{a}_{int}(\omega) = \frac{1}{(\gamma_c + \gamma_{int})/2 -i(\omega-\omega_0)}\, \left(
\sqrt{\gamma_c}\,\hat{a} + \sqrt{\gamma_{int}}\,\hat{f}(\omega)
\right).
\ee
Use of the boundary condition \Eq{eqn:Langevin:input:output:BC} then yields the desired relationship between the output field $\hc$ and the input field $\ha$,
\be{soln:c:Langevin}
\hat{c}(\omega) = \sqrt{\gamma_c}\,\hat{a}_{int}(\omega) - \hat{a}(\omega)
= \left(\frac{\gamma_- + i\delta}{\gamma_+ - i\,\delta}\right)\,\hat{a}(\omega)
+ \frac{\sqrt{\gamma_c\,\gamma_{int}}}{\gamma_+ - i\delta} \, \hat{f}(\omega).
\ee
where we have defined $\gamma_\pm = (\gamma_c \pm \gamma_{int})/2$
and $\delta=\omega-\omega_0$.
Note that \Eq{soln:c:Langevin} has the form of,
\be{eqn:c:form}
\hc = \mathcal{A}_{a\to c}\,\ha + \mathcal{B}\,\hf,
\ee
with $|\mA_{a\to c}|^2 + |\mB|^2 = 1$. Since the input $\ha$ and noise field $\hf$ are independent, they commute and this latter condition ensures that $[\hc(\om),\hc^\dagger(\om')]=\delta(\om-\om')$.
The inclusion of loss for the internal ring resonator mode $\haint$ requires the introduction of noise operators $\hf$ to ensure the preservation of quantum commutations relations. This is the essence of the quantum Langevin approach.
Note that without internal loss ($\gint=0$), $|\mA_{a\to c}|=1$ and the output field $\hc$ is just a phase-shifted version of the input field $\ha$ \cite{Hach:2016,Walls_Milburn:1994,Mandel_Wolf:1995,Scully_Zubairy:1997,Orszag:2000}.
%\cite{Collett_Gardiner:1984,Walls_Milburn:1994,Mandel_Wolf:1995,Scully_Zubairy:1997,Orszag:2000}

%==================================
\subsection{Transmission/cross coupling coefficient derivation: classical}\label{sec:rt:deriv:classical}
\begin{figure}[h]
%\begin{tabular}{cc}
%\includegraphics[width=3.0in,height=2.5in]{rr_loss_all_thru_RR_a_c_aint} &
%\includegraphics[width=3.0in,height=2.5in]{rr_losses_all_thru_RR_a_c} % Original Figure name
\includegraphics[width=3.0in,height=2.5in]{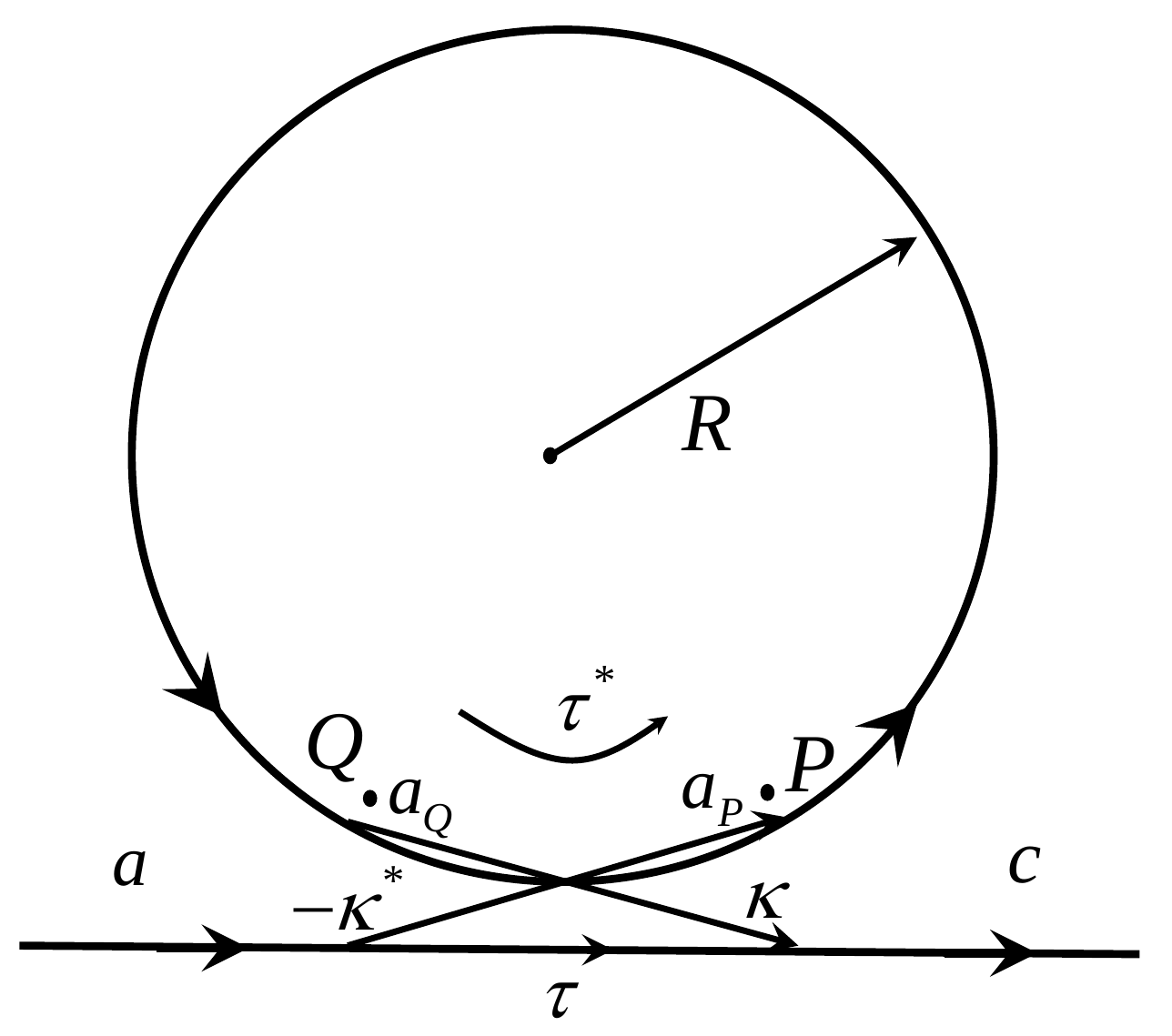}
%\end{tabular}
\caption{An all through ring resonator
}\label{fig:all_thru_RR}
\end{figure}

%==================================
In \Fig{fig:all_thru_RR} we follow the multiple transmission and cross coupling paths in the ring resonator.
We use the notation of \cite{Hach:2014,Preble:2015} in which $\tau$ is the transmission from the input (classical) mode $a$ to output $c$ along the straight waveguide (i.e. $a\to c$) bus and $-\kappa^*$ is the cross coupling
from mode $a$ into the ring resonator (i.e. from $a\to P$). Similarly, $\kappa$ is the cross coupling
\footnote{If the junction of the ring resonator with the external waveguide bus is considered as a beam splitter interaction, the cross coupling coefficients act an as effective reflection coefficients, see \Eq{Rabus:eqn}.}
from inside the ring resonator to the waveguide bus (i.e $Q\to c$) and $\tau^*$ is the internal transmission within the ring (i.e. from $Q\to P$).
The output mode $c$ is obtained as the coherent sum of all possible round trip `Feynman paths' circulating inside the resonator including a round trip amplitude loss $\a=e^{-\frac{1}{2}\G\,L}$ \cite{Yariv:2000}
and phase accumulation $e^{i\theta}$ where $\theta=\beta(\om) L = n(\om)\,\om/c\,L$, with $L=2\pi R$ the perimeter of a ring resonator of radius $R$,
%==================================
%==================================
%===========================
%\bea{Yariv:derivation}
%c = \tau\,a_{a\to c}
%&+& a\,(-\kappa^*)_{a\to P}\,(\alpha\,e^{i\theta})_{P\to Q}\,(\kappa)_{Q\to c}, \no
%%
%&+& a\,(-\kappa^*)_{a\to P}\,(\alpha\,e^{i\theta})_{P\to Q}\,(\tau^*)_{Q\to P}\,
%                             (\alpha\,e^{i\theta})_{P\to Q}\,(\kappa)_{Q\to c}, \no
%%
%&+& a\,(-\kappa^*)_{a\to P}\,(\alpha\,e^{i\theta})_{P\to Q}\,(\tau^*)_{Q\to P}\,
%                             (\alpha\,e^{i\theta})_{P\to Q}\,(\tau^*)_{Q\to P}\,
%                             (\alpha\,e^{i\theta})_{P\to Q}\,(\kappa)_{Q\to c}, \no
%%
%&+& \ldots, \no
%%
%&=& \left( \tau - |\kappa|^2\,\alpha\,e^{i\theta}\,\sum_{n=0}^{\infty} (\tau^*\,\alpha\,e^{i\theta})\right)\,a,\no
%%
%&=&
%\left(
%\frac{\tau - \alpha\,e^{i\theta}}{1-\tau^*\,\alpha\,e^{i\theta}}
%\right)\,a.
%\eea
%which gives the same result as Yariv and Rabus.
%===========================
% align the following in 2 positions, by hand
\begin{subequations}
\bea{Yariv:derivation}
c &=& \tau\,a_{a\to c}
+ a\,(-\kappa^*)_{a\to P}\,(\alpha\,e^{i\theta})_{P\to Q}\,(\kappa)_{Q\to c}, \label{Yariv:derivation:line1} \\
&{}& \hspace{0.45in} +\;  a\,(-\kappa^*)_{a\to P}\,(\alpha\,e^{i\theta})_{P\to Q}\,(\tau^*)_{Q\to P}\,
                             (\alpha\,e^{i\theta})_{P\to Q}\,(\kappa)_{Q\to c}, \label{Yariv:derivation:Line2} \\
&{}& \hspace{0.45in} +\; a\,(-\kappa^*)_{a\to P}\,(\alpha\,e^{i\theta})_{P\to Q}\,(\tau^*)_{Q\to P}\,
                             (\alpha\,e^{i\theta})_{P\to Q}\,(\tau^*)_{Q\to P}\,
                             (\alpha\,e^{i\theta})_{P\to Q}\,(\kappa)_{Q\to c}, \label{Yariv:derivation:Line3} \\
&{}& \hspace{0.45in} +\; \ldots, \no
&=& \left( \tau - |\kappa|^2\,\alpha\,e^{i\theta}\,\sum_{n=0}^{\infty} (\tau^*\,\alpha\,e^{i\theta})^n\right)\,a, \label{Yariv:derivation:Line4}\\
&=&
\left(
\frac{\tau - \alpha\,e^{i\theta}}{1-\tau^*\,\alpha\,e^{i\theta}}
\right)\,a. \label{Yariv:derivation:line5}
\eea
\end{subequations}
In the above, the first term in \Eq{Yariv:derivation:line1} is the direct transmission of mode $a\to c$ (zero round trips), and
in the last line we have used $|\t|^2 + |\k|^2=1$, which states conservation of energy/power.
The notation used in the second term of \Eq{Yariv:derivation:line1} indicates the factors picked up by the mode $a$ as it undergoes one round trip in the resonator, namely $(-\kappa^*)_{a\to P}$ as it cross couples with strength $-\kappa^*$ from the external bus to a point $P$ just inside the ring,  $(\alpha\,e^{i\theta})_{P\to Q}$ as it circulates once around the cavity from the point $P$ to the point $Q$ just before exiting the ring where it out couples $(\kappa)_{Q\to c}$ with strength $\kappa$ to the external mode $c$. \Eq{Yariv:derivation:Line2} and \Eq{Yariv:derivation:Line3} explicitly track two and three circulations respectively around the ring resonator. The sum of all possible circulations is given in \Eq{Yariv:derivation:Line4} which reduces to the final expression \Eq{Yariv:derivation:line5},
which is the classical result as derived
in \cite{Yariv:2000,Heebner_Grover_Ibrahim:2000,Rabus:2007}.

%==================================
\subsection{Conventional matrix `beam splitter' derivation}
%==================================
The derivation in the previous section
is equivalent to the matrix `beam splitter' formulation of Rabus \cite{Rabus:2007},
with $\tau,\,\tau^*$ acting as  transmission coefficients and $\kappa,\,-\kappa^*$ acting as a `reflection' coefficients
between the input modes $a$ and $a_Q$ and output modes $c$ and $a_P$,
\begin{subequations}
\bea{Rabus:eqn}
\left(
  \begin{array}{c}
    c \\
    a_P \\
  \end{array}
\right) &=&
\left(
  \begin{array}{cc}
    \tau & \kappa \\
    -\kappa^* & \tau^* \\
  \end{array}
\right)
\left(
  \begin{array}{c}
    a \\
    a_Q \\
  \end{array}
\right), \\
a_Q &=&  \alpha\,e^{i\theta} \, a_P \label{Rabus:eqn:BC}.
\eea
\end{subequations}
Here, $a_P$ can be considered as the (classical) field cross coupled from the input mode $a$ to just inside the ring resonator at the point $P$. The mode $a_Q$ is the field $a_P$ propagated around the ring once, which suffers a roundtrip loss  $\alpha\equiv e^{-\frac{1}{2}\,\Gamma L}$, with combined coupling and internal loss $\Gamma$ and ring circumference $L=2\pi R$, and a single roundtrip phase accumulation of $\theta$. By solving for $a_P$ from \Eq{Rabus:eqn} and using the internal round trip boundary condition \Eq{Rabus:eqn:BC} we obtain the solution,
\begin{subequations}
\be{Rabus:eqn:a_P}
a_P = \frac{-\kappa^*}{1-\tau^*\,\alpha\,e^{i\theta}}
\ee
which upon using the boundary condition \Eq{Rabus:eqn:BC} yields,
\be{Rabus:eqn:a_Q}
a_Q = \frac{-\kappa^*\,\alpha\,e^{i\theta}}{1-\tau^*\,\alpha\,e^{i\theta}}.
\ee
\end{subequations}
Finally, the first equation in \Eq{Rabus:eqn} $c=\tau\,a + \kappa\,a_Q$ yields the same solution as in \Eq{Yariv:derivation:line5}.

%==================================
%\newpage
\section{Quantum transmission/cross coupling coefficient derivation of output field(s) of a  ring resonator}\label{sec:pma:rr:deriv}
%==================================
\subsection{Quantum derivation}\label{subsec:quantum:all_thru:deriv}
For the quantum derivation, we use the expression \Eq{aL} in \App{app:loudon} (see \Fig{fig:loudon_bs_loss}) for the attenuation loss of a traveling wave, modeled from a continuous set of beams splitters acting as scattering centers due to Loudon
\cite{,Loudon:1997,Loudon:2000},
\be{aL:body}
\hat{a}_L(\omega) = e^{i\xi(\omega)L} \, \hat{a}_0(\omega) + i \sqrt{\Gamma(\omega)}\,\int_{0}^{L} dz \,e^{i\xi(\omega)(L-z)}\,\hat{s}(z,\omega),
\ee
where for convenience we have introduced the shorthand
notation for the input field at $z=0\,$,  $\hat{a}_0(\omega)\equiv\hat{a}(z,\omega)|_{z=0}$ and
the output field at $z=L\,$, $\hat{a}_L(\omega)=\hat{a}(L,\omega)$.
In \Eq{aL:body} we have defined the complex propagation constant as
$\xi(\omega) \equiv \beta(\omega) + i \Gamma(\omega)/2$, with $\beta(\omega) \equiv n(\omega) (\omega/c)$ for a
medium of index of refraction $n(\omega)$ and attenuation constant $\Gamma(\omega)$.
Note that since $\hat{s}(z,\omega)$ are input noise operators, and $\hat{a}_0(\omega)$ is the input field before any interactions with the scattering centers, these operators commute,
\be{comm:rels:a:s}
[\hat{a}_0(\omega),\hat{s}(z',\omega')]= [\hat{a}_0(\omega),\hat{s}^\dagger(z',\omega')]= 0.
\ee
with commutation relations,
\be{comm:rels:a0:s:cont}
[\hat{a}_0(\omega),\hat{a}_0^\dagger(\omega')] = \delta(\omega-\omega'), \quad
[\hat{s}(z,\omega),\hat{s}^\dagger(z',\omega')] = \delta(z-z')\,\delta(\omega-\omega').
\ee
Thus, if we explicitly form the commutation relation $[\hat{a}_L(\omega), \hat{a}_L^\dagger(\omega')]$ we obtain two terms,
\bea{comm:rels:aL}
[\hat{a}_L(\omega), \hat{a}_L^\dagger(\omega')] &=&
e^{i[\xi(\omega)-\xi^*(\omega')]L} \, [\hat{a}_0(\omega),\hat{a}_0^\dagger(\omega')] \no
&+&  \sqrt{\Gamma(\omega)\Gamma(\omega')}\,\int_{0}^{L} dz \int_{0}^{L} dz'\,e^{i[\xi(\omega)(L-z)-\xi^*(\omega')(L-z')]}\,[\hat{s}(z,\omega), \hat{s}^\dagger(z',\omega')],\no
&=& \delta(\omega-\omega')
\big(\,
e^{-\Gamma(\omega) L} + \Gamma(\omega)\,\int_{0}^{L} dz \,e^{-\Gamma(\omega) z}
\,\big), \no
&=& \delta(\omega-\omega'),
\eea
where in the second equality we have used $i[\xi(\omega)-\xi^*(\omega')] = -\Gamma(\omega)$ and the
commutation relations for $a_0(\omega)$ and $s(z,\omega)$ in \Eq{comm:rels:a0:s:cont}, and that the integral in the second to last
line yields $(1-e^{-\Gamma(\omega) L})/\Gamma$. Thus, the expression for the attenuated traveling wave $\hat{a}_L(\omega)$ in \Eq{aL:body} explicitly preserves the output field commutation relations.

In analogy with the classical field derivation in \Sec{sec:rt:deriv:classical}, we track the operator input field $\ha\equiv\ha_0$ as it couples into the ring resonator cavity making an arbitrary number of circulations around the cavity before it couples out to the output mode $\hc$ (see \Fig{fig:all_thru_RR}),
\bsub
\bea{pma:deriv}
\hc &=& \tau\,\ha_0 \label{pma:deriv:1} \\
&+& (-\ks)_{a\to P}(\ha_0 \stackrel{P\to Q}{\longrightarrow} \ha_1)(\k)_{Q\to c},  \label{pma:deriv:2} \\
&+& (-\ks)_{a\to P}
(
\ha_0 \stackrel{P\to Q}{\longrightarrow} \ha_1
       \stackrel{Q\to P}{\longrightarrow} \ts\ha_1
       \stackrel{P\to Q}{\longrightarrow} \ts\ha_2
)
(\k)_{Q\to c},  \label{pma:deriv:3} \\
&+& (-\ks)_{a\to P}
(
\ha_0 \stackrel{P\to Q}{\longrightarrow} \ha_1
      \stackrel{Q\to P}{\longrightarrow} \ts\ha_1
      \stackrel{P\to Q}{\longrightarrow} \ts\ha_2
      \stackrel{Q\to P}{\longrightarrow} \ts^2\ha_2
      \stackrel{P\to Q}{\longrightarrow} \ts^2\ha_3
)
(\k)_{Q\to c},  \label{pma:deriv:4} \\
&+& \ldots,\no
&=& \t\,\ha_0 - |\k|^2\,\sum_{n=0}^\infty (\ts)^n\,\ha_{n+1}, \label{pma:deriv:5} \\
&=& \big( \t - |\k|^2\,\sum_{n=0}^\infty (\ts\alpha e^{i\theta})^n\big)\,\ha_0
-i |\k|^2\,\sqrt{\Gamma}\sum_{n=0}^\infty (\ts)^n\,\Int{(n+1)},  \label{pma:deriv:6} \\
&=&
\left(
\frac{\t-\alpha\,e^{i\theta}}{1-\ts\,\alpha\,e^{i\theta}}
\right)\,\ha
-i |\k|^2\,\sqrt{\Gamma}\sum_{n=0}^\infty (\ts)^n\,\Int{(n+1)}. \label{pma:deriv:7}
\eea
\esub
In \Eq{pma:deriv:1} we have the direct transmission of the input mode
$\ha_0\equiv\ha$ into the output mode $\hc$, while in
\Eq{pma:deriv:2}- \Eq{pma:deriv:4} we follow the round trip evolution of the internal ring resonator mode with $\ha_{n}\equiv\ha_{nL}$ after $n$ round trips through the cavity.
In \Eq{pma:deriv:6} we have used the definition,
\be{anplus1:defn}
\ha_{n+1} \equiv \ha\big((n+1)L,\om\big)= e^{i\xi(\om) (n+1) L}\ha_0(\om) + \Int{(n+1)}
\ee
with
%$\xi(\om) = \beta(\om) + i \Gamma(\om)/2$ such that
$e^{i\xi L}  \equiv \alpha\,e^{i\theta}$ with  $\alpha =e^{-\frac{1}{2}\Gamma L}$
and $\theta = \beta L$.
The above notation is meant to similar to \Eq{Yariv:derivation:line1} with the added annotation
$\ha_0 \stackrel{P\to Q}{\longrightarrow} \ha_1$ indicating that the operator mode $\ha_0$ is transformed into the operator mode $\ha_1$ after one internal circulation within the ring from point $P$ to point $Q$.
The notation $\ha_1\stackrel{Q\to P}{\longrightarrow} \ts\ha_1$ indicates that the mode $\ha_1$ picks up a factor $\ts$ as it internally transmits from the point $Q$ to the point $P$ for the start of an additional circulation within the ring (as opposed to out coupling with strength $(\k)_{Q\to c}$ from the ring resonator at point $Q$
into the external bus mode $\hc$).

As derived in \App{app:pma:comm:deriv} an explicit calculation of the output field commutation relation yields,
\be{c:comm:pma}
[\hc(\om),\hc^\dagger(\om')]=\delta(\om-\om').
\ee
The coefficient of the first term in \Eq{pma:deriv:7} \cite{Yariv:2000} is identical in form to classical transmission coefficient in \Eq{Yariv:derivation:line5}, while the second operator term in \Eq{pma:deriv:7} is the Langevin noise term required to preserve the commutation relation \Eq{c:comm:pma}. Note that in \Eq{pma:deriv:7} we assumed without loss of generality, a single uniform propagation wavevector $\beta(\om)$ and loss $\Gamma(\om)$ throughout the ring resonator. As shown in \App{app:pma:comm:deriv} this assumption can be relaxed and the commutation relations \Eq{c:comm:pma} still hold for multiple, piecewise defined propagation wavevectors and losses along the ring resonator of perimeter length $L$.

%==================================
\subsection{Comparison with quantum Langevin approach}
We now wish to compare the two expressions for the transmission amplitude $\mA_{a\to c}$ from the input mode $\ha$ to the output mode $\hc$ in the single bus ring resonator given by \Eq{soln:c:Langevin}
for the Langevin approach and by \Eq{pma:deriv:7} for the OVPA approach. The power transfer from  $\ha$ to $\hc$ is given by $P_{a\to c} = |\mA_{a\to c}|^2/T_R$, where $T_R = L/v_g$ is the round trip time in the ring resonator of perimeter $L=2\pi R$,
and $v_g$ is the group velocity within the ring.
%and $v_g = (d\beta/d\omega)^{-1}$ is the group velocity within the ring.

For the Langevin case \Eq{soln:c:Langevin}, the expression for
$P_{a\to c}^{(Langevin)}\,T_R = (\gm^2 +\d^2)/(\gp^2 + \d^2)$
has validity around a single resonance at frequency $\om_0$
(see appendices \ref{app:haus} and \ref{app:walls_milburn}).
By construction, the expression using \Eq{pma:deriv:7} for
$P_{a\to c}^{(OVPA)}\,T_R = |(|\t|-\a\,e^{i\theta'})/(1-|\t|\,\a \,e^{i\theta'})|^2$
for the `reflection/transmission' derivation
(defining $\t=|\t|\,e^{i\theta_\t}$ and total phase $\theta'=\theta-\theta_\t$)
 is valid for all resonances as a function of
$\theta=\beta(\om)\,L = \om\,T_R$. Thus, in a neighborhood of a particular resonance at frequency $\om_0$ we have
$\Delta\theta' = T_R\,\d$ with $\d=\om-\om_0$ for which we approximate
$\cos\Delta\theta'\approx 1-\Delta\theta'^2/2= 1-T_R^2\,\d^2/2$.
Substituting this approximation into $P_{a\to c}^{(OVPA)}\,T_R$, keeping terms to order $\d^2$, and equating this to $P_{a\to c}^{(Langevin)}\,T_R$ yields,
\be{eqn:powers}
P_{a\to c}\,T_R =
\frac{\displaystyle\frac{(\a-|\t|)^2}{\a\,|\t|\,T_R^2}+\d^2}{\displaystyle\frac{(1-\a\,|\t|)^2}{\a\,|\t|\,T_R^2}+\d^2}
=
\frac{\gm^2 + \d^2}{\gp^2 + \d^2},
\ee
from which we can read off the expressions,
\be{eqns:gp:gm:compare}
\gp\,T_R = \frac{1-\a\,|\t|}{\sqrt{\a\,|\t|}}, \qquad
\gm\,T_R = \frac{\a-|\t|}{\sqrt{\a\,|\t|}},
\ee
or equivalently,
\be{eqns:gp:gm:compare}
\gc\,T_R   = \frac{(1+\a)\,(1-|\t|)}{\sqrt{\a\,|\t|}}, \qquad
\gint\,T_R = \frac{(1-\a)\,(1+|\t|)}{\sqrt{\a\,|\t|}}
\ee
where we recall that $\a = e^{-\frac{1}{2}\G\,L}$. The expressions in \Eq{eqns:gp:gm:compare} are consistent in the limit of
zero coupling and internal losses $\gc=0$ and $\gint=0$ respectively, i.e. $\G=0$, which yields $\a=|\t|=1$. Following \cite{Heebner_Grover_Ibrahim:2000} we can define a distributed loss for the OVPA case as,
\be{G:Gprime:defn}
|\t| \equiv e^{-\G_{\t}\,L/2}, \qquad \a \equiv e^{-\G\,L/2},
\ee
In the limit of weak losses, we can expand these exponentials to first order in $\G\,L$ and $\G_\t\,L$
and substitute into \Eq{eqns:gp:gm:compare} to obtain,
\be{losses:comparison}
\gc\,T_R \approx \G_{\t}\,L, \qquad \gint\,T_R \approx \G\,L.
\ee
Thus, in the OVPA approach, the magnitude of the transmission coefficient $|\t|$ for power flowing from mode $\ha$ to $\hc$ represents a distributed loss at rate $\gc$, the cavity decay rate, and round trip ring loss $\a$ represents a distributed internal loss at the rate $\G=\gint$. In general, the $\G$ in \Eq{losses:comparison} is frequency dependent and are applicable in the proximity of each resonance $\d=\om-\om_0=0$.

%==================================
%\newpage
\subsection{Add/Drop ring resonator}\label{subsec:add_drop_rr}
\begin{figure}[h]
%\begin{tabular}{cc}
%\includegraphics[width=3.0in,height=2.5in]{rr_losses_all_thru_RR_a_c_aint} &
%\includegraphics[width=3.0in,height=2.75in]{rr_losses_add_drop_RR_a_c_b_d} % Original Figure Name
\includegraphics[width=3.0in,height=2.75in]{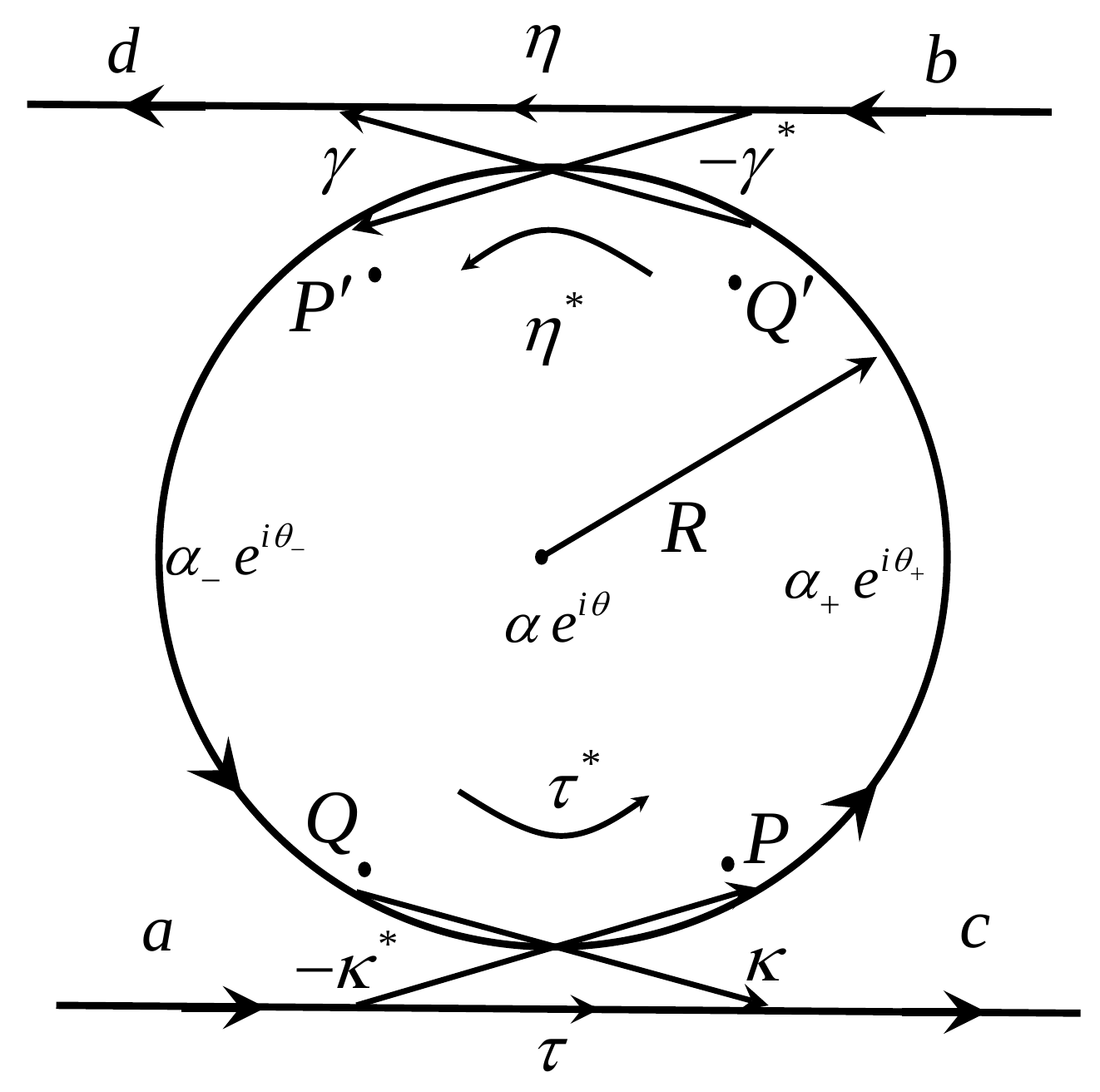}
%\end{tabular}
\caption{An add/drop ring resonator
}\label{fig:add_drop_RR}
\end{figure}

%==================================
We can extend the formalism of the previous section to  consider the quantum derivation of the input-output relations for an add/drop ring resonator as illustrated in \Fig{fig:add_drop_RR}.
%==================================
%==================================
Here $b$ is the (classical) mode injected at the \textit{add port} and
$d$ is mode emitted at the \textit{drop port}. We label as $P'$ the point just inside the ring resonator at which $b$ enters the cavity, and similarly $Q'$  as the point just before the exit to the external mode $d$. We now divide the internal losses and phase shifts into two half-ring portions via $\a_+\,e^{i\theta_+}$ from $P\to Q'$ and
$\a_-\,e^{i\theta_-}$ from $P'\to Q$ such that $\a=\a_+\,\a_-$ and $\theta=\theta_+ + \theta_-$.

Let us first consider the output mode $c$ of the form,
\be{c:form}
c = \mA_{a\to c}\,a + \mA_{b\to c}\,b,
\ee
generalizing \Eq{Yariv:derivation:line5} for the case of the all through (single bus) ring resonator.
Comparison of \Fig{fig:all_thru_RR} and \Fig{fig:add_drop_RR} as well as  \Eq{Yariv:derivation} shows that
the classical loss and phase accumulation factor $\a\,e^{i\theta}$ is replaced by
$\a\,e^{i\theta}\to (\a_+\,e^{i\theta_+})\,(\eta^*)\,(\a_-\,e^{i\theta_-})=\a\,e^{i\theta}\,\eta^*$
in the single bus amplitude $\mA_{a\to c}$ in \Eq{Yariv:derivation:line5}.

Correspondingly, in the quantum derivation we have $\ha_{n+1}\to (\eta^*)^{n+1}\ha_{n+1}$
in \Eq{pma:deriv:5} such that the contribution to $\hc$ from the input port mode $\ha$ in \Eq{c:form} is
given by $\t\,\ha_0 - |\k|^2\,\eta^*\,\sum_{n=0}^\infty (\ts\,\eta^*)^n\,\ha_{n+1}$ where
$\ha_{n+1}$ is given by  \Eq{anplus1:defn}.

For the add port we have classically,
\begin{subequations}
\bea{Yariv:derivation:add:drop}
\mA_{b\to c}
&=& (-\g^*)_{b\to P'}\,(\a_-\,e^{i\theta_-})_{P'\to Q}\,(\kappa)_{Q\to c} \label{Yariv:derivation:add:drop:line1} \\
&+&  (-\g^*)_{b\to P'}\,(\a_-\,e^{i\theta_-})_{P'\to Q}\,(\tau^*)_{Q\to P}\,
                             (\a_+\,e^{i\theta_+})_{P\to Q'}\,(\eta^*)_{Q'\to P'}\,
                              (\a_-\,e^{i\theta_-})_{P'\to Q}\,(\kappa)_{Q\to c}\qquad\label{Yariv:derivation:line2} \\
%
%&{}& \hspace{0.45in} +\; a\,(-\kappa^*)_{a\to P}\,(\alpha\,e^{i\theta})_{P\to Q}\,(\tau^*)_{Q\to P}\,
%                             (\alpha\,e^{i\theta})_{P\to Q}\,(\tau^*)_{Q\to P}\,
%                             (\alpha\,e^{i\theta})_{P\to Q}\,(\kappa)_{Q\to c}, \label{Yariv:derivation:line3} \\
%%
&+& \ldots, \no
&=&
-\g^*\,\kappa\,\a\,e^{i\theta/2}\,\sum_{n=0}^{\infty} (\tau^*\eta^*\,\,\alpha\,e^{i\theta})^n, \label{Yariv:derivation:line3}\\
&=&
-\frac{\g^*\,\kappa\,\a\,e^{i\theta/2}}{1-\tau^*\eta^*\,\alpha\,e^{i\theta}}. \label{Yariv:derivation:line4},
\eea
\end{subequations}
where in \Eq{Yariv:derivation:add:drop:line1} the internal mode picks up a `half-circulation' loss
$\a_-\,e^{i\theta_-}=\sqrt{\a}\,e^{i\theta/2}$ \cite{Rabus:2007}
in traveling from the insertion point $P'$ to the exit point $Q$ a distance $L/2$ away \footnote{Without loss of generality and for algebraic simplicity we have assumed that loss and phase accumulation in each half-circulation of the ring resonator are identical, $\alpha_+=\alpha_- =\sqrt{\alpha}$ and $\theta_+=\theta_- = \theta/2$. These are not a crucial assumptions.
\Eq{aLLprime} and \Eq{comm:rels:aLLprime} show that one can assume an arbitrary number of different piecewise constant losses along the lengths $L_i$ of the ring such that $\sum_i\,L_i=L$. Similar considerations hold for the phase accumulation.}.
In the quantum derivation, this corresponds to a
contribution in \Eq{c:form} to $\hc$ from the add port mode $\hb$
given by $-\g^*\,\k\,\sum_{n=0}^\infty (\ts\,\eta^*)^n\,\hb_{n+1/2}$. Here
$\hb_{n+1/2}$ is given by an analogous expression in \Eq{anplus1:defn} with $\ha\to\hb$ and $n+1\to n+1/2$,
corresponding to the classical `half-circulation' loss.
Thus, \Eq{c:form} takes the form (with $\ha_0\equiv \ha$ and $\hb_0\equiv\hb$ indicating modes just inside the ring resonator experiencing zero round trips),
\bsub
\bea{c:form:2}
c &=& \mA_{a\to c}\,a + \mA_{b\to c}\,b,\no
\Rightarrow \hc &=&
\t\,\ha_0 - |\k|^2\,\eta^*\,\sum_{n=0}^\infty (\ts\,\eta^*)^n\,\ha_{n+1}
-\g^*\,\k\,\sum_{n=0}^\infty (\ts\,\eta^*)^n\,\hb_{n+1/2}, \\
&=&
\left(
\frac{\t-\eta^*\,\a\,e^{i\theta}}{1-\t\,\eta^*\,\a\,e^{i\theta}}
\right)\,\ha
-
\left(
\frac{\g^*\,\k\,\sqrt{\a}\,e^{i\theta/2}}{1-\t\,\eta^*\,\a\,e^{i\theta}}
\right)\,\hb
-i\,\sqrt{\G}
\left(
|\k|^2\,\eta^*\hf_{a} + \g^*\,\k \hf_{b}
\right),
\eea
\esub
where we have define the noise operators as,
\be{noise:fa:fb}
\hf_{a} = \sum_{n=0}^{\infty} (\ts\eta^*)^n\,\hat{s}_{n+1},\quad
\hf_{b} = \sum_{n=0}^{\infty} (\ts\eta^*)^n\,\hat{s}_{n+1/2},\quad
\hat{s}_{m}=\int_{0}^{m L} dz\,e^{i\xi(\om)[m L - z]}\,\hat{s}(z,\om).
\ee

A similar analysis can be carried out for the drop port mode $\hd$ in terms of the input $\ha$ and add port $\hb$ modes, yielding,
\bsub
\bea{d:form}
d &=& \mA_{a\to d}\,a + \mA_{b\to d}\,b,\no
\Rightarrow \hd &=&
-
\left(
\frac{\k^*\,\g\,\sqrt{\a}\,e^{i\theta/2}}{1-\t\,\eta^*\,\a\,e^{i\theta}}
\right)\,\ha
+\left(
\frac{\eta-\ts\,\a\,e^{i\theta}}{1-\t\,\eta^*\,\a\,e^{i\theta}}
\right)\,\hb
-i\,\sqrt{\G}
\left(
\k^*\,\g \hf_{a} + |\g|^2\,\ts\hf_{b}
\right).
\eea
\esub
Note, for the zero loss case $\a=1$ the transition amplitudes $\mA_{a\to c}$, $\mA_{b\to c}$, $\mA_{a\to d}$, $\mA_{b\to d}$ are the same ones derived classically in \cite{Rabus:2007} and quantum mechanically in \cite{Hach:2014} for the add/drop ring resonator.
The preservation of the commutation relations
$[\hc(\om),\hc^\dagger(\om')]=[\hd(\om),\hd^\dagger(\om')]=\delta(\om-\om')$ and
$[\hc(\om),\hd(\om')]=[\hc(\om),\hd^\dagger(\om')]=0$
can be explicitly demonstrated straightforwardly (though with somewhat more involved algebra)
 through the approach used in \App{app:pma:comm:deriv} for
explicitly proving the all through commutation relation \Eq{c:comm:pma}

\section{Hong-Ou-Mandel Manifold with loss}\label{sec:HOMM_with_loss}
\begin{figure}[h]
%\begin{tabular}{cc}
%\includegraphics[width=3.0in,height=2.5in]{rr_losses_all_thru_RR_a_c_aint} &
%\includegraphics[width=3.0in,height=2.75in]{rr_losses_add_drop_RR_a_c_b_d} % Original Figure Name
%\includegraphics[width=3.0in,height=2.75in]{P11_leq_0p001_alpha_1p0_coffee_can_mac_red}
\includegraphics[width=3.0in,height=2.75in]{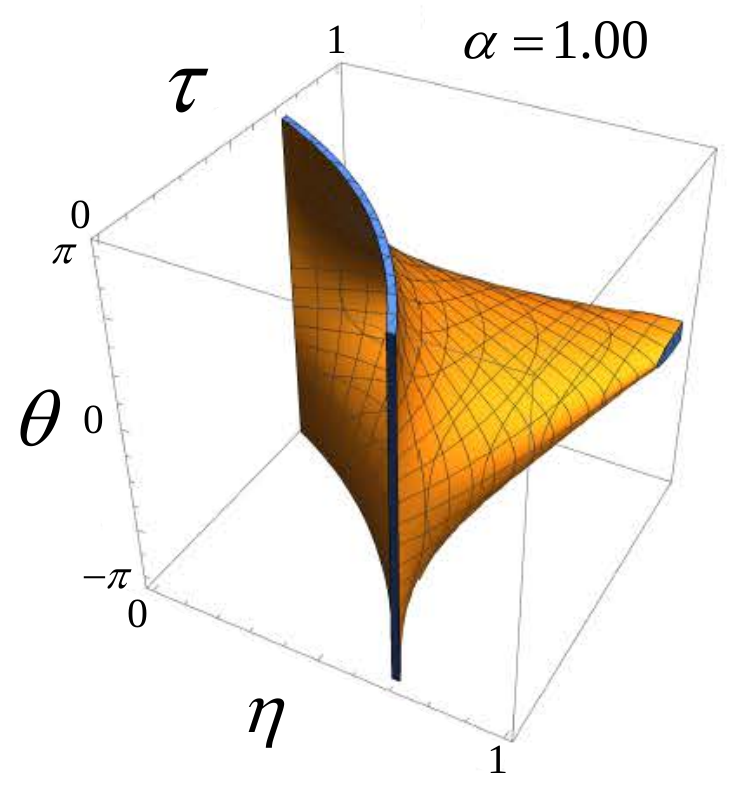}
%\end{tabular}
\caption{(Color online) Hong-Ou-Mandel manifold (HOMM) for $P_{c,d}(1,1) \le 0.001$ for zero loss $\alpha = 1$, as a function of through-coupling parameters $0\le\tau\le 1$ and $0\le\eta\le 1$ for the system output modes $c$ and $d$ respectively of \Fig{fig:add_drop_RR}, and the internal single round trip phase accumulation $-\pi\le\theta\le \pi$ (compare with Fig.(5b) of \cite{Hach:2014}).
}\label{fig:P11_leq_0p001_alpha_1}
\end{figure}

In this section we re-examine the Hong-Ou-Mandel manifold (HOMM) introduced by Hach {et. al.} \cite{Hach:2014}
for the lossless add/drop double bus ring resonator in the previous \Sec{subsec:add_drop_rr}, but now
using the expressions for the output modes $c$ \Eq{c:form:2} and and $d$ \Eq{d:form} which includes the effects of internal and coupling losses.
The HOMM is defined by the level surface $P_{c,d}(1,1)=0$ for the destructive interference of the coincident output photon state $\ket{1_c, 1_d}$
(given the input state $\ket{1_a, 1_b}$ ) containing one photon in each system output mode $c$ and $d$ (see \Fig{fig:add_drop_RR}) as a function of the through-coupling parameters $\tau$ and $\eta$ (for modes $c$ and $d$ respectively), and the internal single round trip phase accumulation $\theta$. In \Fig{fig:P11_leq_0p001_alpha_1} we plot the region $0\leq P_{c,d}(1,1)\leq 0.001$ corresponding to $99.9\%$ destructive interference \cite{epsilon:note} of the quantum amplitude for the state $\ket{1_c, 1_d}$ for the real parameters $0\leq \tau,\eta \leq 1$ (with the cross-coupling parameters giving by $\kappa = \sqrt{1-\tau^2}$ and $\gamma = \sqrt{1-\eta^2}$) and $-\pi\leq \theta \leq  \pi$.
%==================================
%==================================
As discussed in Hach {et. al.} \cite{Hach:2014}, the two dimensional (three parameter) HOMM arising in the lossless add/drop ring resonator generalizes the zero dimensional (one parameter) Hong-Ou-Mandel effect \cite{HOM:1987} where the single adjustable parameter is the transmissivity of the 50:50 beam splitter upon which the two photons interfere.

To examine the effects of coupling and intrinsic loss on the HOMM in the add/drop ring resonator, we begin with the input state
$\ket{1_a, 1_b,0_{env}}\equiv\ket{1_a, 1_b}\otimes\ket{0}_{env}$ where $\ket{0}_{env}$ represents the (for simplicity, zero temperature) initial vacuum state of the noise modes which are acted upon by the noise operators $\hat{f}_a$ and $\hat{f}_b$  defined in \Eq{noise:fa:fb}. Let us write \Eq{c:form:2} and \Eq{d:form} for the output modes $c$ and $d$ in terms of the system input modes $a$ and $b$ formally as
\be{out_modes}
\hat{\vec{c}}_{out} =
\left(
  \begin{array}{c}
    \hat{c} \\
    \hat{d} \\
  \end{array}
\right) =
\left(
  \begin{array}{cc}
    \mathcal{A}_{a\rightarrow c} & \mathcal{A}_{b\rightarrow c} \\
    \mathcal{A}_{a\rightarrow d} & \mathcal{A}_{b\rightarrow d} \\
  \end{array}
\right) \,
\left(
  \begin{array}{c}
    \hat{a} \\
    \hat{b} \\
  \end{array}
\right) +
\left(
  \begin{array}{c}
    \hat{F}_c \\
    \hat{F}_d \\
  \end{array}
\right)
\equiv M \, \hat{\vec{a}}_{in} + \hat{\vec{F}},
\ee
where we have defined the collective noise operators
$\hat{F}_c = -i\,\sqrt{\G}\,\left(|\k|^2\,\eta^*\hf_{a} + \g^*\,\k \hf_{b}\right)$,
and
$\hat{F}_d = -i\,\sqrt{\G}\,\left(\k^*\,\g \hf_{a} + |\g|^2\,\ts\hf_{b}\right)$.
From the definition \Eq{noise:fa:fb} we see that $\hat{f}_a$ depends on an integer number of round trip losses in the ring resonator
(i.e. mode $\hat{a}\rightarrow\hat{c}$ or $\hat{b}\rightarrow\hat{d}$ involving the noise operator $\hat{s}_{n+1}(z,\omega)$),
while $\hat{f}_b$ depends on an integer plus half number of round trip losses
(i.e. mode $\hat{a}\rightarrow\hat{d}$ or $\hat{b}\rightarrow\hat{c}$ involving the noise operator $\hat{s}_{n+1/2}(z,\omega)$).
Thus, while $[\hat{f}_a,\hat{f}_b]=0$, we have $[\hat{f}_a,\hat{f}^\dagger_b]\ne0$. This is to be expected \cite{Barnett:1996,Agarwal:2014} due to the feedback (sum over multiple round trips) provided by the ring resonator. While the commutator $[\hat{f}_a,\hat{f}^\dagger_b]$ could be explicitly computed directly as in \Sec{app:pma:comm:deriv} (for the single bus ring resonator) we can now invoke (as is typically done) the unitarity of the input modes and output modes commutators to determine the value of the noise commutators.
Returning to \Eq{out_modes} in terms of the collective noise modes $\hat{F}_c$ and $\hat{F}_d$  we can infer that
\bsub % need {} before "[" due to "weird error" in texbook
\bea{FcFd:comm}
[\hat{c}(\om),\hat{c}^\dagger(\om')] = \delta(\om-\om')
     & \Rightarrow & [\hat{F}_c(\om),\hat{F}_c^\dagger(\om')] = \left(1 - (|\mathcal{A}_{a\rightarrow c}|^2+|\mathcal{A}_{b\rightarrow c}|^2)\right)\, \delta(\om-\om'), \label{FcFcdag:comm}\\
{} [\hat{d}(\om),\hat{d}^\dagger(\om')] = \delta(\om-\om')
     &\Rightarrow& [\hat{F}_d(\om),\hat{F}_d^\dagger(\om')] = \left(1 - (|\mathcal{A}_{a\rightarrow d}|^2+|\mathcal{A}_{b\rightarrow d}|^2)\right)\, \delta(\om-\om'), \label{FdFddag:comm}\\
{} [\hat{c}(\om),\hat{d}^\dagger(\om')] = 0
     &\Rightarrow& [\hat{F}_c(\om),\hat{F}_d^\dagger(\om')] =
     -\left( \mathcal{A}_{a\rightarrow c}\,\mathcal{A}^*_{a\rightarrow d}\, + \mathcal{A}_{b\rightarrow c}\,\mathcal{A}^*_{a\rightarrow d} \right)\, \delta(\om-\om'). \label{FcFddag:comm}
\eea
\esub

The input state $\ket{\Psi}_{in} = \ket{1_a, 1_b,0_{env}} = \ha^\dagger \hb^\dagger \ket{0_a, 0_b,0_{env}}$ is converted to the
output state $\ket{\Psi}_{out}$ by rewriting the input modes operators $\hat{a}^\dagger$ and $\hat{b}^\dagger$
in terms of the output mode operators $\hat{c}^\dagger$ and $\hat{d}^\dagger$.
Inverting \Eq{out_modes} as
\be{in_modes}
 \hat{\vec{a}}^\dagger_{in} = {\mathcal{M}} \, (\hat{\vec{c}}^\dagger_{out} - \hat{\vec{F}}^\dagger), \qquad {\mathcal{M}} = M^{-1 *},
\ee
yields the output state
\be{Psi_out}
\ket{\Psi}_{out} \equiv
\ket{\Psi^{(2)}}_{c,d}\otimes\ket{0}_{env}
+ \ket{\phi^{(1)}}_{c,d}\otimes\hat{F}_c^\dagger\ket{0}_{env}
+ \ket{\varphi^{(1)}}_{c,d}\otimes\hat{F}_d^\dagger\ket{0}_{env}
+  \ket{0,0}_{c,d}\otimes\ket{\Phi^{(2)}}_{env},
\ee
where
%\bsub
%\bea{Psi_out:terms}
%\ket{\Psi^{(2)}}_{c,d} &=&
%\sqrt{2}\,\M_{1 1}\,\M_{2 1}\ket{0,2}_{c,d} + (\M_{1 1}\,\M_{2 2} + \M_{1 2}\,\M_{2 1} )\ket{1,1}_{c,d} + \sqrt{2}\,\M_{1 2}\,\M_{2 2}\ket{2,0}_{c,d}, \quad \\
%%
%\ket{\phi^{(1)}}_{c,d} &=& -\left( 2\,\M_{1 1}\,\M_{2 1}\ket{1,0}_{c,d} + (\M_{1 1}\,\M_{2 2} + \M_{1 2}\,\M_{2 1}) \ket{0,1}_{c,d} \right), \\
%%
%\ket{\phi^{(1)}}_{c,d} &=& -\left(  (\M_{1 1}\,\M_{2 2} + \M_{1 2}\,\M_{2 1}) \ket{0,1}_{c,d} + 2\,\M_{1 2}\,\M_{2 2}\ket{1,0}_{c,d} + \right), \\
%%
%\ket{\Phi^{(2)}}_{env} &=& (\M_{1 1} \hat{F}_c^\dagger + \M_{1 2} \hat{F}_d^\dagger)\, (\M_{1 1} \hat{F}_c^\dagger + \M_{1 2} \hat{F}_d^\dagger)\,\ket{0}_{env}.
%\eea
%\esub
%%
\bsub
\bea{Psi_out:terms}
\ket{\Psi^{(2)}}_{c,d} &=&
\sqrt{2}\,\M_{1 1}\,\M_{2 1}\ket{2,0}_{c,d} + \textrm{Perm}(\M)\,\ket{1,1}_{c,d} + \sqrt{2}\,\M_{1 2}\,\M_{2 2}\ket{0,2}_{c,d}, \quad \label{Psi_out:terms:2}\\
\ket{\phi^{(1)}}_{c,d} &=& -\left( 2\,\M_{1 1}\,\M_{2 1}\ket{1,0}_{c,d} + \textrm{Perm}(\M)\, \ket{0,1}_{c,d} \right), \label{Psi_out:terms:1a}\\
\ket{\varphi^{(1)}}_{c,d} &=& -\left(  \textrm{Perm}(\M)\, \ket{1,0}_{c,d} + 2\,\M_{1 2}\,\M_{2 2}\ket{0,1}_{c,d} \right), \label{Psi_out:terms:1b}\\
\ket{\Phi^{(2)}}_{env} &=& (\M_{1 1} \hat{F}_c^\dagger + \M_{1 2} \hat{F}_d^\dagger)\, (\M_{2 1} \hat{F}_c^\dagger + \M_{2 2} \hat{F}_d^\dagger)\,\ket{0}_{env}, \label{Psi_out:terms:0}
\eea
\esub
where we have defined
\be{PerM}
\textrm{Perm}(\M) \equiv \M_{1 1}\,\M_{2 2} + \M_{1 2}\,\M_{2 1},
\ee
as the permanent \cite{Scheel:2008,*Scheel:2004} of the matrix $\M$.

Ultimately we are interested in the observable reduced system density matrix $\rho_{c,d} = \textrm{Tr}_{env}[\ket{\Psi}_{out}\langle \Psi|]$ of the output modes $c$ and $d$.
The trace over the environment is facilitated by the observation that e.g.
$\textrm{Tr}_{env}[\hat{F}_i^\dagger\ket{0}_{env}\langle 0| \hat{F}_j]
={}_{env}\bra{0} \hat{F}_j\, \hat{F}_i^\dagger \ket{0}_{env}
={}_{env}\bra{0} [\hat{F}_j, \hat{F}_i^\dagger] + \hat{F}_i^\dagger\,\hat{F}_j  \ket{0}_{env}
= [\hat{F}_j, \hat{F}_i^\dagger]$ for $i,j\in\{c, d\}$ and where use of \Eq{FcFcdag:comm}, \Eq{FdFddag:comm}, and \Eq{FcFddag:comm} can be made.

The reduced system density matrix has the form
\be{rho_cd}
\rho_{c,d} = \sum_{k=\{0,1,2\}} \, p_k\, \rho^{(k)}_{c,d}, \qquad \textrm{Tr}_{c,d}[\rho^{(k)}_{c,d}] = 1,  \qquad \sum_{k=\{0,1,2\}} \, p_k = 1,
\ee
where the index $k$ labels the number of photons in the modes $c$ and $d$.
The 2-system-photon sector $\rho^{(2)}_{c,d}$ is spanned by the states $\{\ket{2,0}_{c,d}, \ket{1,1}_{c,d}, \ket{0,2}_{c,d}\}$,
the 1-system-photon sector $\rho^{(1)}_{c,d}$ is spanned by the states $\{\ket{1,0}_{c,d}, \ket{0,1}_{c,d}\}$, and the
the 0-system-photon sector $\rho^{(0)}_{c,d}$ is the vacuum state $\ket{0}_{c,d}\bra{0} $.

%==================================
\begin{figure}[ht]
%\begin{tabular}{cc}
%\includegraphics[width=3.0in,height=2.5in]{rr_losses_all_thru_RR_a_c_aint} &
%\includegraphics[width=3.0in,height=2.75in]{rr_losses_add_drop_RR_a_c_b_d} % Original Figure Name
%\includegraphics[width=6.0in,height=3.5in]{P11_leq_0p001_PltPts_300_6Feb2017_vary_alpha_red}
\includegraphics[width=6.0in,height=3.75in]{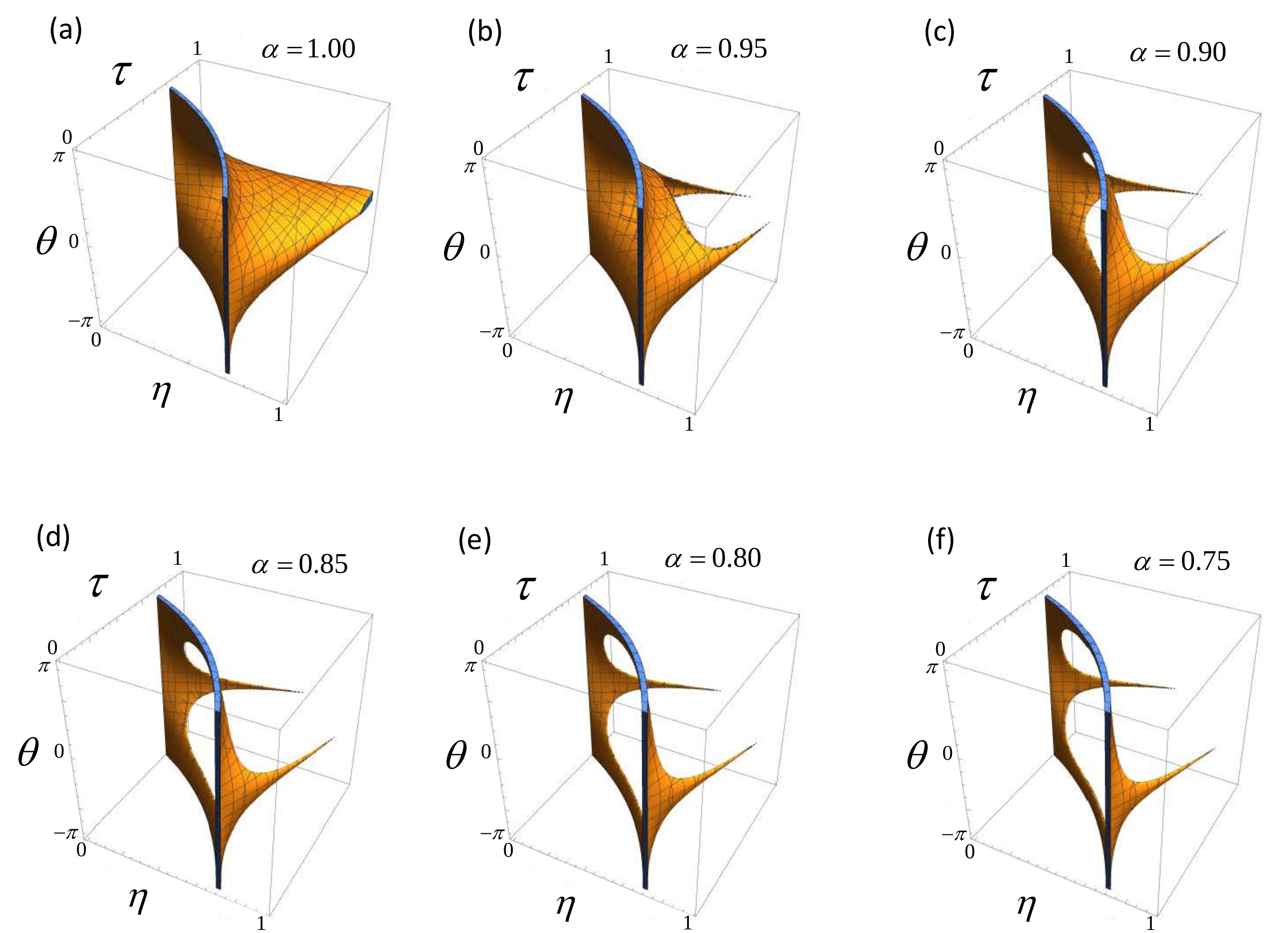}
%\end{tabular}
\caption{(Color online) Hong-Ou-Mandel manifold (HOMM) for $0\le P^{(\alpha)}_{c,d}(1,1) \le 0.001$ as a function of the loss parameter $\alpha = e^{-\G L/2}$, the through-coupling parameters $0\le\tau\le 1$ and $0\le\eta\le 1$ for the system output modes $c$ and $d$ respectively of \Fig{fig:add_drop_RR}, and the internal single round trip phase accumulation $-\pi\le\theta\le \pi$.
(a) $\alpha=1.0$ (lossless), (b) $\alpha=0.95$, (c) $\alpha=0.90$, (d) $\alpha=0.85$, (e) $\alpha=0.80$, (f) $\alpha=0.75$ (compare with \Fig{fig:P11_leq_0p001_alpha_1}).
}\label{P11_le_0p001_vary_alpha}
\end{figure}
%==================================
Finally, $P_{c,d}^{(\alpha)}(1,1) \equiv {}_{c,d}\bra{1,1} \rho^{(1)}_{c,d} \ket{1,1}_{c,d}$ is the probability, as function of the loss parameter $\alpha = e^{-\G L/2}$, that a coincidence detection will contain  one output photon in mode $c$ and one output photon in mode $d$ for the diagonal density matrix $\rho^{(1)}_{c,d}$. (Such events occur randomly with probability $p_2$).
From \Eq{Psi_out:terms:2} we see that $P_{c,d}^{(\alpha)}(1,1)=\textrm{Perm}(\M)$ as has been recently noted in the theory of
generalized multiphoton (i.e HOM) quantum interference effects, especially in regards to the problem of boson sampling  \cite{Sanders:2013,*Sanders:2014,*Sanders:2015,*Tichy:2015}.
%==================================
\begin{figure}[ht]
%\begin{tabular}{cc}
%\includegraphics[width=3.0in,height=1.75in]{P11_theta_alpha_1-0p75} &
%\includegraphics[width=3.0in,height=1.75in]{P11_theta_alpha_0p75-0p50} % Original Figure Name
%\end{tabular}
%\includegraphics[width=5.0in,height=2.0in]{critical_coupling_plot}
\includegraphics[width=6.0in,height=2.10in]{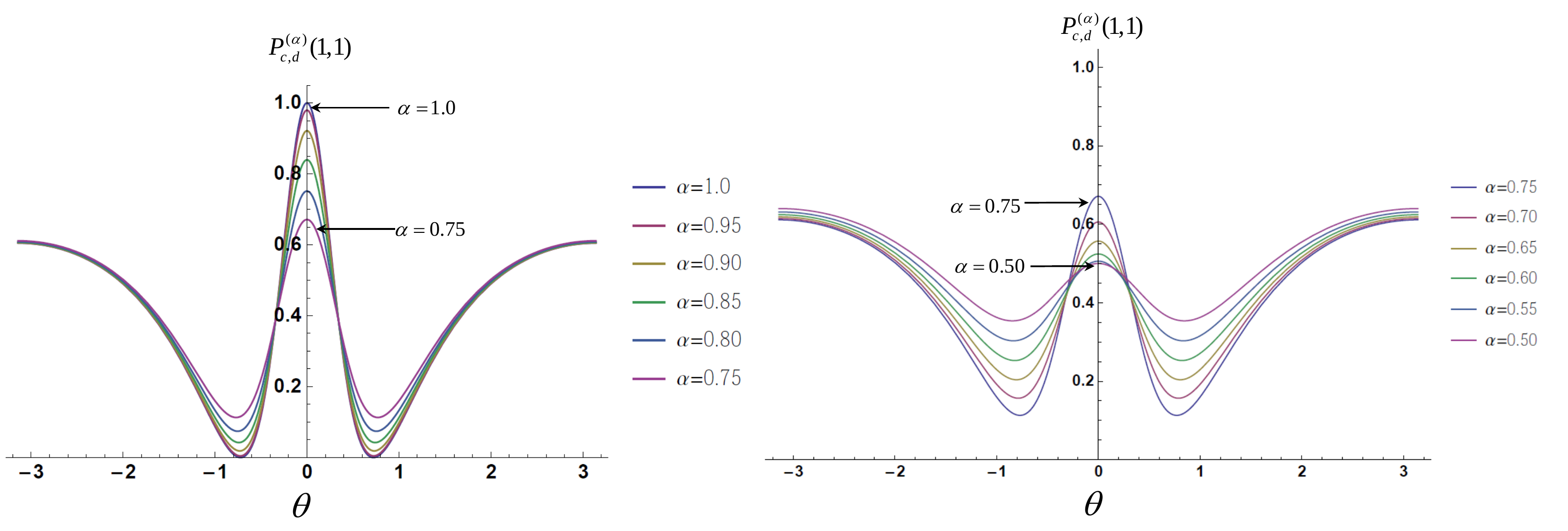}
\caption{(Color online) Hong-Ou-Mandel manifold (HOMM) for $P^{(\alpha)}_{c,d}(1,1) =0$ for the important special case of critical coupling $\tau=\eta=1/\sqrt{2}$ (i.e. 3dB couplers) versus the internal single round trip phase accumulation $-\pi\le\theta\le \pi$ for various loss parameters $\alpha = e^{-\G L/2}$.
(left) $0.75\le\alpha\le 1.0$, (right) $0.50\le\alpha\le 0.75$ (compare Fig.(6b) of \cite{Hach:2014}).
}\label{P11_tau_eq_eta_critical_coupling}
\end{figure}
%==================================
The expression for $P_{c,d}^{(\alpha)}(1,1)$ is given by
\bea{P11:alpha}
&{}& P_{c,d}^{(\alpha)}(1,1) =\no
&{}&
\frac{
(|\tau|^2 + \alpha^2\,|\eta|^2 - \alpha\,r)\,(|\eta|^2 + \alpha^2\,|\tau|^2 - \alpha\,r)
+ \alpha^2\,|\kappa|^4\,|\gamma|^4
+ \alpha\,|\kappa|^2\,|\gamma|^2 \, \left((1+\alpha^2)\,r - 2\alpha(|\tau|^2 + |\eta|^2)\right)
}
{
(|\tau|^2 + \alpha^2\,|\eta|^2 - \alpha\,r)\,(|\eta|^2 + \alpha^2\,|\tau|^2 - \alpha\,r)
+ \alpha^2\,|\kappa|^4\,|\gamma|^4
- \alpha\,|\kappa|^2\,|\gamma|^2 \, \left((1+\alpha^2)\,r - 2\alpha(|\tau|^2 + |\eta|^2)\right)
},
\qquad
\eea
where we have defined $r\equiv 2\,\textrm{Re}(\tau\eta\,e^{-i\theta})$.
\Eq{P11:alpha} reduces in the lossless case $\alpha=1$ to
\be{P11:alpha_eq_1}
P_{c,d}^{(\alpha=1)}(1,1) =
\left(
\frac
{|\tau|^2 + |\eta|^2 - r - |\kappa|^2\,|\gamma|^2}
{|\tau|^2 + |\eta|^2 - r + |\kappa|^2\,|\gamma|^2}
\right)^2
\ee
whose numerator (set equal to zero) was examined in \cite{Hach:2014} for the case of the lossless HOMM.

In \Fig{P11_le_0p001_vary_alpha} we plot the region $0\leq P_{c,d}(1,1)\leq 0.001$ corresponding to $99.9\%$ \cite{epsilon:note} destructive interference of the quantum amplitude for the state $\ket{1_c, 1_d}$ for the real parameters $0\leq \tau,\eta \leq 1$ (with the cross-coupling parameters giving by $\kappa = \sqrt{1-\tau^2}$ and $\gamma = \sqrt{1-\eta^2}$) and $-\pi\leq \theta \leq \pi$. The HOMM begins to break up at approximately $5\%$ loss ($\alpha=0.95$), and reduces to essentially a lower dimension manifold for loss greater than $10\%$ ($\alpha<0.90$). Currently, loss in silicon ring resonators at $1550$nm can be as low as $1\%$ \cite{Hach:2016,Preble:2015} so that the observation of the HOMM appears experimentally feasible.

In \Fig{P11_tau_eq_eta_critical_coupling} we plot $P^{(\alpha)}_{c,d}(1,1) =0$ for the important special case of critical coupling $\tau=\eta=1/\sqrt{2}$ (i.e. 3dB couplers) versus the internal single round trip phase accumulation $\theta$ for various loss parameters $0.5\le\alpha\le 1.0$. As the internal and coupling loss ($\G$) increases
($\alpha = e^{-\G L/2}$ decreases) we observe the expected disappearance of the HOM dip (zero minima for the lossless case $\alpha=1.0$) and the decrease in visibility (difference between maximum value at $\theta=0$ and minimum values of $P^{(\alpha)}_{c,d}(1,1) =0$). Again, we can see that for up to $5\%$ loss ($0.95\le\alpha\le 1.0$) the observation of the HOMM appears experimentally feasible.

%=======================================================
% 1-sys-photon sector
%=======================================================
It is also interesting to examine the one system photon sector $\rho^{(1)}_{c,d}$ of the reduced density matrix $\rho_{c,d}$
spanned by the basis states $\{\ket{1,0}_{c,d}, \ket{0,1}_{c,d}\}$. Let us define the un-normalized state $\tilde{\rho}^{(1)}_{c,d}$ as
\be{rho_tilde_1}
\tilde{\rho}^{(1)}_{c,d}
=\ket{\phi^{(1)}}_{c,d}\bra{\phi^{(1)}}\,[\hat{F}_c,\hat{F}_c^\dagger]
+\ket{\phi^{(1)}}_{c,d}\bra{\varphi^{(2)}}\,[\hat{F}_d,\hat{F}_c^\dagger]
+\ket{\varphi^{(2)}}_{c,d}\bra{\phi^{(1)}}\,[\hat{F}_c,\hat{F}_d^\dagger]
+\ket{\varphi^{(2)}}_{c,d}\bra{\varphi^{(2)}}\,[\hat{F}_d,\hat{F}_d^\dagger],\qquad
\ee
and $p_1 = \textrm{Tr}[\tilde{\rho}^{(1)}_{c,d}]$, then $\rho^{(1)}_{c,d} = \tilde{\rho}^{(1)}_{c,d}/p_1$.
Note that $\rho^{(1)}_{c,d}=\textrm{Tr}[ \ket{\Psi^{(1)}}_{out}\,\bra{\Psi^{(1)}} ]$
arises from the trace over the environment of the (post-selected) one system photon portion of $\ket{\Psi_{out}}$ in \Eq{Psi_out} where
\be{Psi_1_out}
\ket{\Psi^{(1)}}_{out} \equiv \frac{1}{\sqrt{p_1}}\,
\left(
\ket{\phi^{(1)}}_{c,d}\otimes\hat{F}_c^\dagger\ket{0}_{env}
+ \ket{\varphi^{(1)}}_{c,d}\otimes\hat{F}_d^\dagger\ket{0}_{env}
\right),
\ee
and hence $\ket{\Psi^{(1)}}_{out}$ could be considered as the (system-environment) purification of the
(post-selected, with probability $p_1$) system state $\rho^{(1)}_{c,d}$.
As such, the entropy $S^{(1)} = -\textrm{Tr}[\,\rho^{(1)}_{c,d}\,\log_2\,\rho^{(1)}_{c,d}\,]$ indicates a measure of the bipartite entanglement between the system and environment for the post-selected state $\ket{\Psi^{(1)}}_{out}$.
%==================================
\begin{figure}[ht]
%\begin{tabular}{cc}
%\includegraphics[width=3.0in,height=1.75in]{P11_theta_alpha_1-0p75} &
%\includegraphics[width=3.0in,height=1.75in]{P11_theta_alpha_0p75-0p50} % Original Figure Name
%\end{tabular}
%\includegraphics[width=5.0in,height=4.0in]{S_plot_1-sys-photon}
\includegraphics[width=5.0in,height=4.0in]{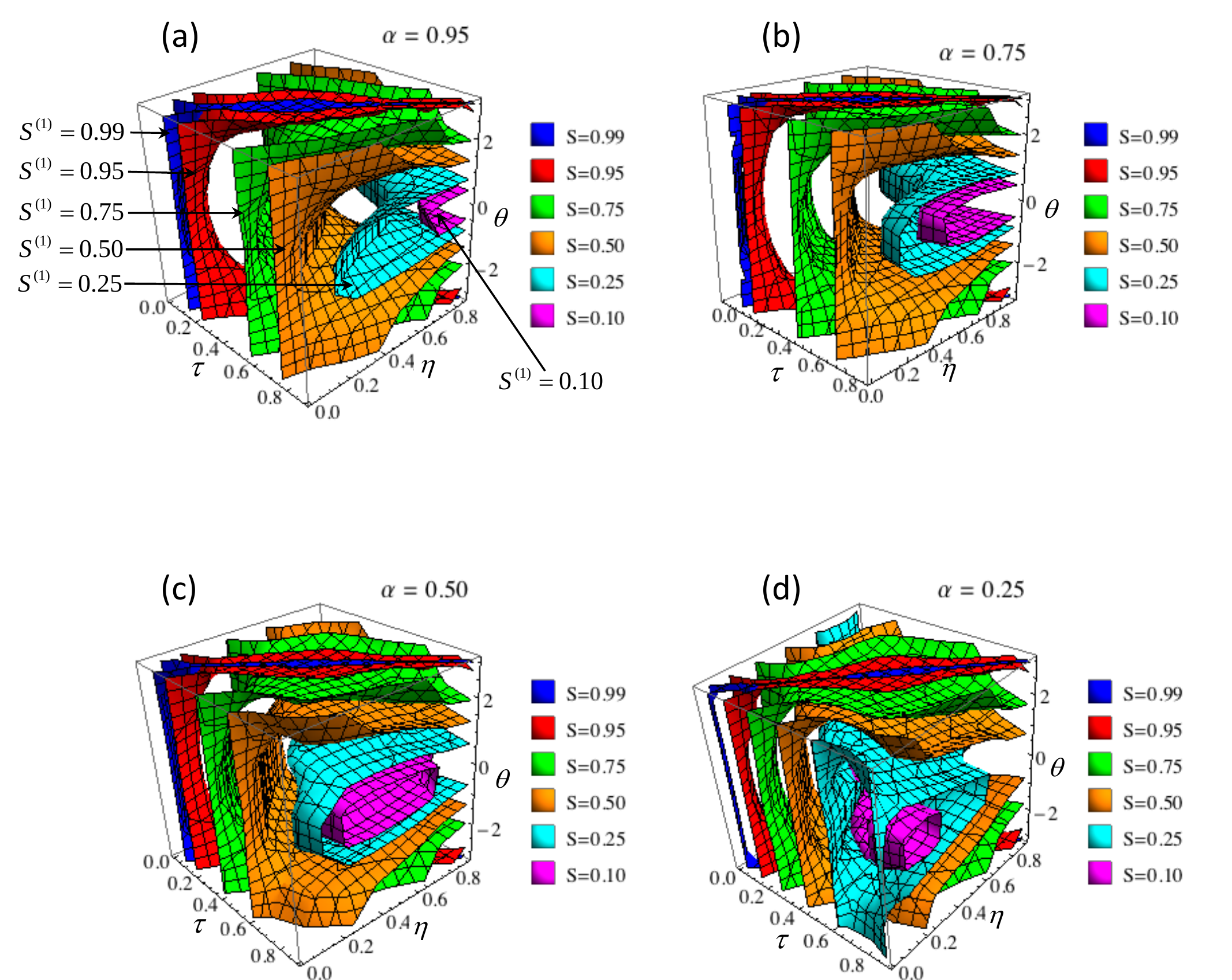}
\caption{
(Color online) Contour plots of the von Neumann entropy
$S^{(1)} = -\textrm{Tr}[\,\rho^{(1)}_{c,d}\,\log_2\,\rho^{(1)}_{c,d}\,]$
for the one system photon sector of $\rho_{c,d}$ as a function of through-coupling parameters
$0\le\tau\le 1$ and $0\le\eta\le 1$,
and the internal single round trip phase accumulation $-\pi\le\theta\le \pi$,
for various loss parameters $\alpha = e^{-\G L/2}$.
(a) $\alpha=0.95$, (b) $\alpha=0.75$, (c) $\alpha=0.50$, (d) $\alpha=0.25$.
(Surface manifolds for contour values of $S^{(1)} \in\{0.99, 0.95, 0.75, 0.50, 0.25, 0.10\}$ retain the same 
nested relative  orientation (from left to right) for all subplots $(a)-(d)$  as that labeled in $(a)$).
}
\label{S_plot_1_sys_photon}
\end{figure}
%==================================
In \Fig{S_plot_1_sys_photon} we plot level surfaces of $S^{(1)}$ as a function of $\tau$, $\eta$ and $\theta$ for various values of the loss parameter $\alpha$.
Values of $S^{(1)}$ closer to unity indicate greater entanglement between single system photon (in mode $c$ and $d$), and the single photon lost to the environment in the post-selected state $\ket{\Psi^{(1)}}_{out}$. These regions of larger entanglement are diminished as loss is increased ($\alpha$ decreased).

Lastly, it is interesting to note that from the definition of
$\ket{\phi^{(1)}}_{c,d}$ in \Eq{Psi_out:terms:1a} and
$\ket{\varphi^{(1)}}_{c,d}$ in \Eq{Psi_out:terms:1b}
that both states are suppositions of the one system photon basis states  $\{\ket{1,0}_{c,d}, \ket{0,1}_{c,d}\}$.
These superpositions are completely destroyed precisely at the condition that HOMM is strongest, namely
$P^{(\alpha)}_{c,d}(1,1)= \textrm{Perm}(\M)=0$.

%=======================================================
\section{Summary and Outlook}\label{conclusion}
%\vspace{-1em}\flushleft{({\textit{\color{red} Conclusion and future work outlook still needed})}
In this paper we have examined quantum optical losses in ring resonators using field operator transformations. Specifically, we have demonstrated the equivalence between our operator valued phasor addition of `Feynman paths' circulating within the resonator and the more standard Langevin approach. In fact, we have shown that the OVPA approach we present here is slightly more general in that it is valid for all frequencies of light while the Langevin only holds near a resonance of the system. This result represents an important `unification' of  the description of such networks based upon scattering theory with that based upon quantum transfer functions (matrices). With the results of this paper in place, we can now investigate the quantum optical response of ring resonator networks to exotic states of light in the presence of losses
%and under conditions of finite bandwidths.
We will apply the techniques developed here and elsewhere in the references to design and optimize silicon nanophotonic networks for quantum information processing, optical metrology, and communication.

Note, after the completion of this work, the authors were made aware of the paper by
Raymer and McKinstrie (2013) \cite{Raymer:2013} which considered a generalization of the standard Langevin input-output formalism that explicitly  takes into account circulation factors accounting for the multiple round trips of the fields inside a cavity or ring resonator. That work considered an equation of motion for one round trip of a single bus cavity field with no internal losses, along with auxiliary beam-splitter like boundary conditions relating the input and output fields to the circulating cavity field. While not explicitly including internal propagation losses, the authors indicated how they would be included in a Langevin approach.
The current work discussed in this paper is similar in spirit, but considers directly the
total summation of all round trip circulations of the field(s) in a \textit{lossy} (coupling and propagation) single bus and dual bus ring resonator without the use of boundary conditions. The two approaches are equivalent to each other. Both works consider the agreement of the formalism with the standard Langevin approach in the high cavity Q limit.
%=======================================================
%\newpage
\appendix
%=======================================================
\section{Classical derivation of input-output fields}\label{app:haus}
In the interest of making this paper as self-contained as possible, we review in this appendix the classical derivation of the input-output formalism by Haus \cite{Haus:1984,Haus:1997,Haus:2000}, relating the coupling of an internal cavity (complex) amplitude $a_{int}$ to an external input $a_{in}$ and output field $a_{out}$ as illustrated in \Fig{fig:haus_one_sided_cavity}.
Since the optical system considered here is linear, the classical equations will also hold  in the quantum regime, as will be reviewed in the next appendix, where consideration of commutation relations must be additionally taken into account. The phenomenological derivation by Haus relies on three principles (i) energy conservation, (ii) time reversibility and (iii) perturbation theory to formulate a dynamical, and boundary condition relation between the internal cavity and the external driving and out-coupled modes.

\subsection{A single cavity resonance}
\begin{figure}[h]
%\begin{tabular}{cc}
%\includegraphics[width=2.25in,height=1.5in]{rr_losses_haus_input_output_one_sided_cavity} %& Original Figure Name
%\includegraphics[width=2.25in,height=1.5in]{fig4} %&
\includegraphics[width=2.25in,height=1.5in]{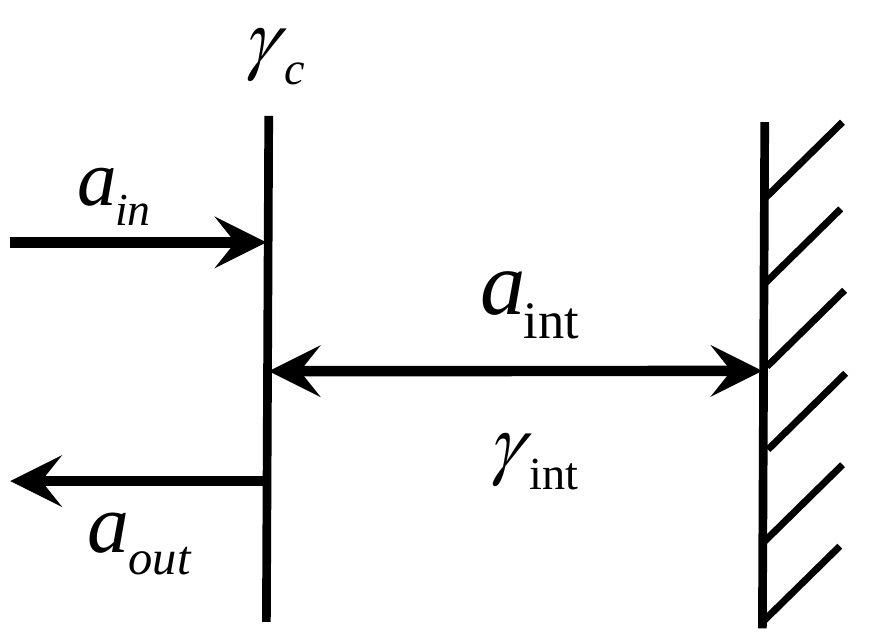} %&
%\includegraphics[width=3.0in,height=2.5in]{all_thru_RR_a_c}
%\end{tabular}
\caption{One-sided cavity with classical input field amplitude $a_{in}$, output field $a_{out}$ and internal cavity field
$a_{int}$. $\gamma_c$ is the (power) decay rate of the internal field to the external field through the mirror. $\gamma_{int}$ represents internal losses within the cavity.
}\label{fig:haus_one_sided_cavity}
\end{figure}

The equation of motion for the internal field $a_{int}$ in a one-sided lossy Fabry-Perot cavity, as illustrated in
\Fig{fig:haus_one_sided_cavity}, driven by an external field $a_{in}$ and out-coupled to the external field $a_{out}$ is given by,
%We can now extend the above results to include internal losses inside the cavity which decay at a rate of $\gamma_{int}$. \Eq{aint:loss:driven:2} is modified to
\be{aint:loss:driven:3}
\dot{a}_{int} = -(i\,\omega_0 + \gamma_c/2 + \gamma_{int}/2)\,a_{int} + \sqrt{\gamma_c}\,a_{in}.
\ee
%==================================
%==================================
Here, $\omega_0$ is the resonance frequency of the undriven cavity, $\gamma_{c}$ is the
power decay rate of the internal field through the partial mirror to the external mode $(d|a_{int}|^2/dt = -\gamma_c\,a_{int})$, and $\gamma_{int}$ describes internal (e.g. scattering) losses within the cavity. The term $\sqrt{\gamma_c}\,a_{in}$
describes the in-coupling of the form $\kappa\,a_{in}$ of the external field of complex amplitude $a_{in}$ with coupling constant $\kappa$. One can relate  the coupling constant $\kappa$ to the cavity decay rate $\gamma_c$ through energy conservation and time reversal as $\kappa^2 = \gamma_c$ (for detailed derivation see Haus \cite{Haus:1984,Haus:1997,Haus:2000}).
For a driving field excitation $a_{in}$ proportional to $e^{-i\omega\,t}$, the internal field has the solution,
\be{aint:soln:3}
a_{int} = \frac{\kappa\,a_{in}}{\gamma_c/2+\gamma_{int}/2 -i\,(\omega - \omega_0)}.
\ee
describing a complex Lorentzian form of width $(\gamma_c+ \gamma_{int})/2$.

One can relate the output field $a_{out}$ to the input $a_{in}$ and internal cavity field $a_{int}$ through power conservation (in appropriately normalized units of energy and power),
\be{power:conserv}
|a_{in}|^2 - |a_{out}|^2 =  \frac{d|a_{int}|^2}{dt} = -\gamma_c\,|a_{int}|^2 + \sqrt{\gamma_c}\,(a_{in}\,a^*_{int} + a^*_{in}\,a_{int}).
\ee
Since the system is linear we can write formally the ansatz $a_{out} = c_{in}\,a_{in} + c_{int}\,a_{int},$
%\be{aout:cin_cout}
%a_{out} = c_{in}\,a_{in} + c_{int}\,a_{int},
%\ee
for some complex constants $c_{in}$ and  $c_{int}$.
%
%\be{energy_conserv:1}
%\frac{d|a_{int}|^2}{dt} = -\gamma_c\,|a_{int}|^2 = - |a_{out}|^2, \quad (a_{in}= \gamma_{int}=0),
%\ee
%
From the case of the undriven cavity with no internal losses  ($a_{in}=\gamma_{int}=0$),
energy conservation $d|a_{int}|^2/dt = -\gamma_c\,|a_{int}|^2 = - |a_{out}|^2$ yields
%\Eq{energy_conserv:1}
$a_{out} = \sqrt{\gamma_c}\,a_{int}$ so that $c_{int} = \sqrt{\gamma_c}$.
Substituting $a_{out} = c_{in}\,a_{in} + \sqrt{\gamma_c}\,a_{int}$ into the left hand side of the above ansatz produces
$|a_{in}|^2\,(c_{int}+1) -\gamma_c\,|a_{int}|^2 -  \sqrt{\gamma_c}\,(c_{in}\,a_{in}\,a^*_{int} + c^*_{in}\,a^*_{in}\,a_{int} )$ which on comparison with the right hand side of \Eq{power:conserv} yields the real solution $c_{int} = -1$. Thus, we obtain,
\be{input_output:BC}
a_{out} = -a_{in} + \sqrt{\gamma_c}\,a_{int}, \quad \textrm{or} \quad a_{in} + a_{out} = \sqrt{\gamma_c}\,a_{int},
\ee
which can be considered as a boundary condition for the fields at the lossy mirror.

Using \Eq{aint:soln:3} and the boundary condition \Eq{input_output:BC}  we can calculate the reflection coefficient $r$ as,
\be{refl:coeff}
r = \frac{a_{out}}{a_{in}} = \frac{\sqrt{\gamma_c}\,a_{int}- a_{in}}{a_{in}}
  = \frac{(\gamma_c - \gamma_{int})/2 + i (\omega-\omega_0)}{(\gamma_c + \gamma_{int})/2 - i (\omega-\omega_0)}
\equiv \frac{\gamma_- + i\delta}{\gamma_+ - i\delta},
\ee
where in the last equality we have defined $\gamma_{\pm}= (\gamma_c \pm \gamma_{int})/2$ and $\delta=\omega-\omega_0$, as in the main body.
Note that when the internal losses are zero $\gamma_{int}=0$ one has $|r|=1$, otherwise $|r|<1.$
\Eq{aint:loss:driven:3} and the boundary condition \Eq{input_output:BC}  describe the internal classical field amplitude $a_{int}$ of the resonator near a single resonance and relates it to the input driving field $a_{in}$ and the external traveling wave mode $a_{out}$ that it couples to. Since the systems is linear, these equations also hold  in the quantum regime, as will be shown in the next appendix, where consideration of commutation relations must be taken into account.

\subsection{Extension to internal losses and multiple resonances}
The generalization to multiple resonances is achieved by writing \Eq{aint:loss:driven:3} for each internal cavity mode $a_{int,j}$ near resonance frequency $\omega_{0,j}$, with individual coupling $\gamma_{c,j}$ and internal losses $\gamma_{int,j}$,
\be{aint:loss:driven:gen}
\dot{a}_{int,j} = -(i\,\omega_{0,j} + \gamma_{c,j}/2 + \gamma_{int,j}/2)\,a_{int,j} + \sqrt{\gamma_{c,j}}\,a_{in}.
\ee
The boundary condition \Eq{input_output:BC} generalizes to,
\be{input_output:BC:gen}
a_{out} = c_{in}\,a_{in} + \sum_j\,\sqrt{\gamma_{c,j}}\,a_{int,j}.
\ee
The reflection coefficient similarly generalizes to,
\be{refl:coeff:gen}
r = \frac{a_{out}}{a_{in}} = c_{in} + \sum_j \, \frac{\gamma_{c,j}}{(\gamma_{c,j} + \gamma_{int,j})/2 - i (\omega-\omega_{0,j})},
\equiv c_{in} + \sum_j \, L_j,
\ee
where we have defined the complex Lorentzian $L_j = \gamma_{c,j}/[\gamma_{c,j} + \gamma_{int,j})/2 - i (\omega-\omega_{0,j})]$.
Again, for zero internal losses $\gamma_{int,j}=0$ we must have $|r|^2=1$ which leads to a quadratic equation for $c_{in}$ (taken as real),
\be{cin:eqn:gen}
(c_{in}+1)\,(c_{in}-1) + (c_{in}+1)\,\sum_j\,|L_j|^2 + 2\,\sum_{j\ne k}\,\textrm{Re}(L_j\,L^*_k)=0.
\ee
We see that $c_{in}$ is now a function of $\omega$.
For a single resonance $j=1$ there is only one term in the sum $\sum_j L_j$ and hence the last term in \Eq{cin:eqn:gen} is not present. By inspection, $c_{in}=-1$ in this case.
%For the general case of well separated resonances (i.e. large free spectral range), near a particular resonance $\omega = \omega_j + \Omega$ such that
%$\Omega \ll |\omega_j - \omega_k|$ for $k\ne j$
For the general case, near a particular resonance $\omega = \omega_j + \Omega$ such that
$\Omega,\gamma_{c,j} \ll |\omega_j - \omega_k|$ for $k\ne j$ (i.e. well separated resonances, large free spectral range)
$|L_{k\ne j}| \approx \gamma_{c,k}/|\omega_j - \omega_k| \ll 1$ so that the last cross term in \Eq{cin:eqn:gen} is negligible and only the single $|L_j|^2$ term contributes to the middle sum. Hence, as in the single resonance case \Eq{cin:eqn:gen} becomes approximately $(c_{in}+1)\,(c_{in}-1) + (c_{in}+1)\,|L_j| \approx 0$ with solution
$c_{in}(\omega_j + \Omega)\approx -1$. Thus, near each individual resonance, the single resonance boundary condition \Eq{input_output:BC} holds.

%=======================================================
\section{Quantum derivation of input-output fields}\label{app:walls_milburn}
The quantum derivation of the input-output relations for optical fields in a cavity is attributed to the work of Collett and Gardiner \cite{Collett_Gardiner:1984}.
Here we follow the often cited texts of Walls and Milburn \cite{Walls_Milburn:1994} and of Orszag \cite{Orszag:2000}.
In this formulation a Hamiltonian is prescribed to yield dynamics of the same form given classically in
\Eq{aint:loss:driven:3} due to the linearity of the system. The quantum version of the classical boundary condition \Eq{input_output:BC} arises from the difference between the equations of motion for the noise operators considered in the far past and far future, which couples the internal cavity mode to the external modes of the cavity. The essential new feature of the quantum derivation is the preservation of the commutation relations of all involved operators, which is required by the unitarity of the quantum evolution. While this material is now standard in quantum optics canon, we include it here for completeness, and for comparison to the OVPA coupling derivation used in the main body of the text.
%=======================================================
\begin{figure}[h]
%\begin{tabular}{cc}
%\includegraphics[width=2.25in,height=1.5in]{rr_losses_walls_milburn_input_output_one_sided_cavity} %& Original Fig Name
%\includegraphics[width=2.25in,height=1.5in]{fig5} %&
\includegraphics[width=2.25in,height=1.5in]{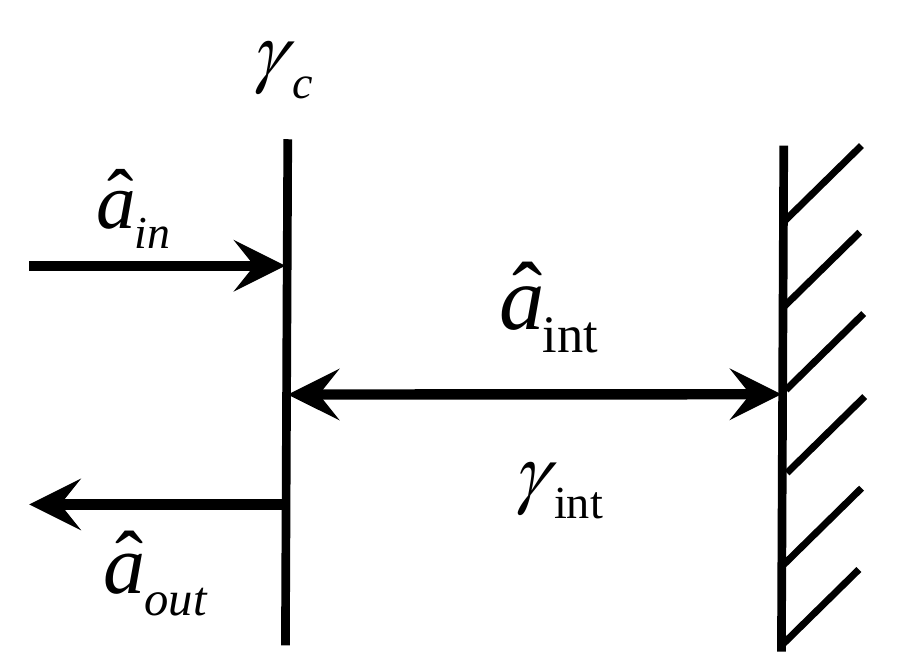} %&
%\includegraphics[width=3.0in,height=2.5in]{all_thru_RR_a_c}
%\end{tabular}
\caption{Same one sided cavity as in \Fig{fig:haus_one_sided_cavity}, except now classical amplitudes have been changed to quantum fields.
}\label{fig:walls_milburn_one_sided_cavity}
\end{figure}
%==================================

The quantum input-output relations are instantiations of the $S$ (scattering) matrix which relates input fields to output fields. Here we assume linear interactions between the system and the bath, the rotating wave approximation and that the spectrum of the bath is flat, independent of frequency. The Hamiltonian is given by,
\bsub
\bea{Hamiltonian:WM}
H &=& H_{sys} + H_B + H_{INT}, \\
H_B &=& \int_{-\infty}^{\infty} d\omega\,\hbar\,\omega\,\hat{b}^\dagger(\omega)\,\hat{b}(\omega), \\
H_{int} &=& i\,\hbar\int_{-\infty}^{\infty} d\omega\, \kappa(\omega)\,\left(\hat{b}^\dagger(\omega)\, \hat{a}_{int} - \hat{b}(\omega)\,\hat{a}^\dagger_{int}\right).
\eea
\esub
Here $\hat{a}_{int}$ is the internal cavity mode, $\hat{b}^\dagger(\omega),\,\hat{b}(\omega)$ are the creation and annihilation operator for the bath modes assumed to have a white noise spectrum such that $[\hat{b}(\omega),\,\hat{b}^\dagger(\omega')]= \delta(\omega-\omega')$, and $\kappa(\omega)$ is the coupling constant. Though the frequencies are positive, the integration range can be extended from $(-\omega_0,\infty)$ in a rotating frame of frequency $\omega_0$. The lower limit of the integral can then be extended to $-\infty$ for $\omega_0\gg\Delta\omega$ where $\Delta\omega$ is the bandwidth of frequencies under consideration (say, near a particular resonance).

The Heisenberg equations of motion yield,
\bsub
\bea{eqns:Heisenberg}
\dot{\hat{b}}(\omega,t) &=& -i\,\omega\,\hat{b}(\omega,t) + \kappa(\omega)\,\hat{a}_{int}, \label{eqn:bdot}\\
\dot{\hat{a}}_{int}(\omega,t) &=& -\frac{i}{\hbar}\,[\hat{a}_{int}, H_{sys}] - \int_{-\infty}^{\infty} d\omega\, \kappa(\omega)\,\hat{a}_{int}. \label{eqn:aintdot}
\eea
\esub
We can solve \Eq{eqn:bdot} for $\hat{b}(\omega,t)$ depending on two different choices of the initial conditions,
\bsub
\bea{solns:b0:b1}
\hat{b}(\omega,t) &=& e^{-i\omega(t-t_0)}\,\hat{b}(\omega,t_0)
             + \int_{t_0}^{t} dt'\, \kappa(\omega)\,e^{-i\omega(t-t')}\,\hat{a}_{int}(t'),  \label{soln:b0}\\
\hat{b}(\omega,t) &=& e^{-i\omega(t-t_1)}\,\hat{b}(\omega,t_1)
             - \int_{t}^{t_1} dt'\, \kappa(\omega)\,e^{-i\omega(t-t')}\,\hat{a}_{int}(t'). \label{soln:b1}
\eea
\esub
In \Eq{soln:b0} the initial condition has been chosen at a time in the far past $t_0<t$ such that $b(\omega,t_0)$ represents the bath operators at very early times (often taken to be $t_0=-\infty$), whereas in \Eq{soln:b1} the initial condition has been chosen to be in the far future $t_1>t$ such that $b(\omega,t_1)$ represents the bath operators at very late times (often taken to be $t_1=\infty$). We also assume that in the far past, the bath and the system are uncorrelated so that the operators commute
$[a_{int},b(\omega,t_0)]=[a_{int},b^\dagger(\omega,t_0)]=0$.

We first consider the substitution of \Eq{soln:b0} into \Eq{eqn:aintdot} to obtain the exact equation,
\bsub
\bea{eqn:aintdot:b0}
\dot{a}_{int}(t) &=& -\frac{i}{\hbar}\,[\hat{a}_{int}, H_{sys}] \no
&-& \int_{-\infty}^{\infty} d\omega\, \kappa(\omega)\,e^{-i\omega(t-t_0)}\,\hat{b}(\omega,t_0), \label{ain:1}\\
&-& \int_{-\infty}^{\infty} d\omega\, \kappa^2(\omega)\,  \int_{t_0}^{t} dt'\, \,e^{-i\omega(t-t_0)}\,\hat{a}_{int}(t'). \label{aint:decay_term}
\eea
\esub
We now invoke the Markov approximation that coupling $\kappa(\omega)$ is constant over the bandwidth $\Delta\omega$ so that we can pull it out from under the integral in term~(\ref{ain:1}). As in \App{app:haus} we relate the coupling constant $\kappa(\omega)$ to the cavity decay rate $\gamma_c$ via $\kappa^2(\omega) = \gamma_c/(2\pi)$.
We further define the remaining integral in term~(\ref{ain:1}) as,
\be{ain:defn}
\hat{a}_{in}(t) \equiv -\frac{1}{\sqrt{2\pi}}\,\int_{-\infty}^{\infty} d\omega\,e^{-i\omega(t-t_0)}\,\hat{b}(\omega,t_0),
\ee
using the sign convention that incoming fields to the cavity have a minus sign, while outgoing fields have a plus sign (see $a_{out}(t)$ below.
Thus, the term~(\ref{ain:1}) becomes $\sqrt{\gamma_c}\,\hat{a}_{in}(t)$.
By use of the definition and properties of the delta function,
%\Eq{delta:defn},
\bsub
\bea{deltafuncs:defns}
\frac{1}{2\pi}\,\int_{-\infty}^{\infty} d\omega\,e^{-i\omega(t-t')} &=& \delta(t-t'), \label{delta:defn}\\
\int_{t_0}^{t} dt'\,f(t')\,\delta(t-t') = \int_{t}^{t_1} dt'\,f(t')\,\delta(t-t') &=& \frac{1}{2}\,f(t), \quad t_0 < t < t_1, \label{delta:property}
\eea
\esub
and the initial bath operator equal time commutation relations $[\hat{b}(\omega,t_0), \hat{b}^\dagger(\omega',t_0)] = \delta(\omega-\omega')$,
one has
\be{ain:comm:rel}
[\hat{a}_{in}(t), \hat{a}^\dagger_{in}(t')] = \delta(t-t').
\ee
By again pulling out $\kappa(\omega) = \gamma_c/(2\pi)$ from under the integral in \Eq{aint:decay_term} and using \Eq{delta:defn} and \Eq{delta:property}
the term in \Eq{aint:decay_term} becomes $-\gamma_c/2\,a_{int}(t)$. Gathering these results together yields the equation for the internal cavity mode $\hat{a}_{int}(t)$,
\be{eqn:aindot:ain:b0}
\dot{\hat{a}}_{int}(t) = -\frac{i}{\hbar}\,[\hat{a}_{int}, H_{sys}] -\frac{\gamma_c}{2}\,\hat{a}_{int}(t) + \sqrt{\gamma_c}\,\hat{a}_{in}(t).
\ee
This is the exact same form as the classical equation of motion for $a_{int}$ in \Eq{aint:loss:driven:3}  if we take
$H_{sys} = \hbar\omega_0\,\hat{a}^\dagger_{int}\,\hat{a}_{int}$ as the free-field, empty cavity Hamiltonian,
and consider no internal losses $\gamma_{int}=0$.
\Eq{eqn:aindot:ain:b0} is the quantum Langevin \cite{Walls_Milburn:1994,Mandel_Wolf:1995,Scully_Zubairy:1997,Orszag:2000}
equation of motion for the internal cavity mode $\hat{a}_{int}$ coupled to the input driving field $\hat{a}_{in}$.
It is an embodiment of the fluctuation-dissipation theorem \cite{Mandel_Wolf:1995} which  states that effect of loss (dissipation) in the system is accompanied by the presence of noise sources (fluctuations) as the cause of the loss. These noise operators must be present quantum mechanically in order to preserve the system commutation relations $[\hat{a}_{int}(t), \hat{a}^\dagger_{int}(t')] = \delta(t-t')$. Otherwise, without the presence of the term $\hat{a}_{in}(t)$ in \Eq{eqn:aindot:ain:b0} the system commutator would decay to zero as $e^{-\gamma_c (t-t')}$.

We can repeat the above development of the equation of motion for $\hat{a}_{int}$, this time using the
solution for $\hat{b}(\omega,t)$ in \Eq{soln:b1} in terms of the far-future modes $\hat{b}(\omega,t_1)$ to obtain,
\be{eqn:aindot:ain:b1}
\dot{\hat{a}}_{int}(t) = -\frac{i}{\hbar}\,[\hat{a}_{int}, H_{sys}] + \frac{\gamma_c}{2}\,\hat{a}_{int}(t) - \sqrt{\gamma_c}\,\hat{a}_{out}(t).
\ee
where we have defined $\hat{a}_{out}(t)$ analogous to \Eq{ain:defn} as,
\be{aout:defn}
\hat{a}_{out}(t) \equiv \frac{1}{\sqrt{2\pi}}\,\int_{-\infty}^{\infty} d\omega\,e^{-i\omega(t-t_1)}\,\hat{b}(\omega,t_1),
\ee
which straightforwardly yields the commutation relations,
\be{aout:comm:rel}
[\hat{a}_{out}(t), \hat{a}^\dagger_{out}(t')] = \delta(t-t'),
\ee
analogous to \Eq{ain:comm:rel}.
Lastly, the boundary condition between the input, output and internal cavity mode is obtained by subtracting the two equations of motion for
$\hat{a}_{int}(t)$ \Eq{eqn:aindot:ain:b0} and \Eq{eqn:aindot:ain:b1} to obtain,
\be{eqn:qBC}
\hat{a}_{in}(t) + \hat{a}_{out}(t) = \sqrt{\gamma_c}\,\hat{a}_{int}(t),
\ee
which has the exact same form as the classical boundary condition obtained in \Eq{input_output:BC}.

Although the above analysis pertains to cavities driven by a bath, it is not necessarily a theory about noise, since the only properties assumed about the bath is flat spectral response \cite{Orszag:2000}. Similar to \Eq{aint:loss:driven:3}, we can explicitly include internal losses, treating $\hat{a}_{in}$ as an external (non-noise) driving field by explicitly including noise operators $\hat{f}(t)$ that are delta correlated in time $[\hat{f}(t),\hat{f}^\dagger(t')]=\delta(t-t')$,
\be{eqn:aindot:ain:f}
\dot{\hat{a}}_{int}(t) = -\frac{i}{\hbar}\,[\hat{a}_{int}, H_{sys}]
                       - \frac{(\gamma_c+\gamma_{int})}{2}\,\hat{a}_{int}(t)
                       + \sqrt{\gamma_c}\,\hat{a}_{in}(t)
                       + \sqrt{\gamma_{int}}\,\hat{f}(t).
\ee

%=======================================================
%\newpage
\section{Loudon's quantum traveling-wave attenuation}\label{app:loudon}
%=======================================================
One of the primary expressions we use in the main body of the paper is Loudon's formulation for traveling-wave attenuation by an infinite series of discrete beam splitters.
Here we summarize Loudon's   derivation \cite{,Loudon:1997,Loudon:2000} and note several important points on the commutation relations for the effective noise operator expressions.

\begin{figure}[h]
%\begin{tabular}{cc}
%\includegraphics[width=6.0in,height=2.25in]{rr_losses_loudon_bs_loss} %& Original Figure Name
%\includegraphics[width=6.0in,height=2.25in]{fig6} %&
\includegraphics[width=6.0in,height=2.25in]{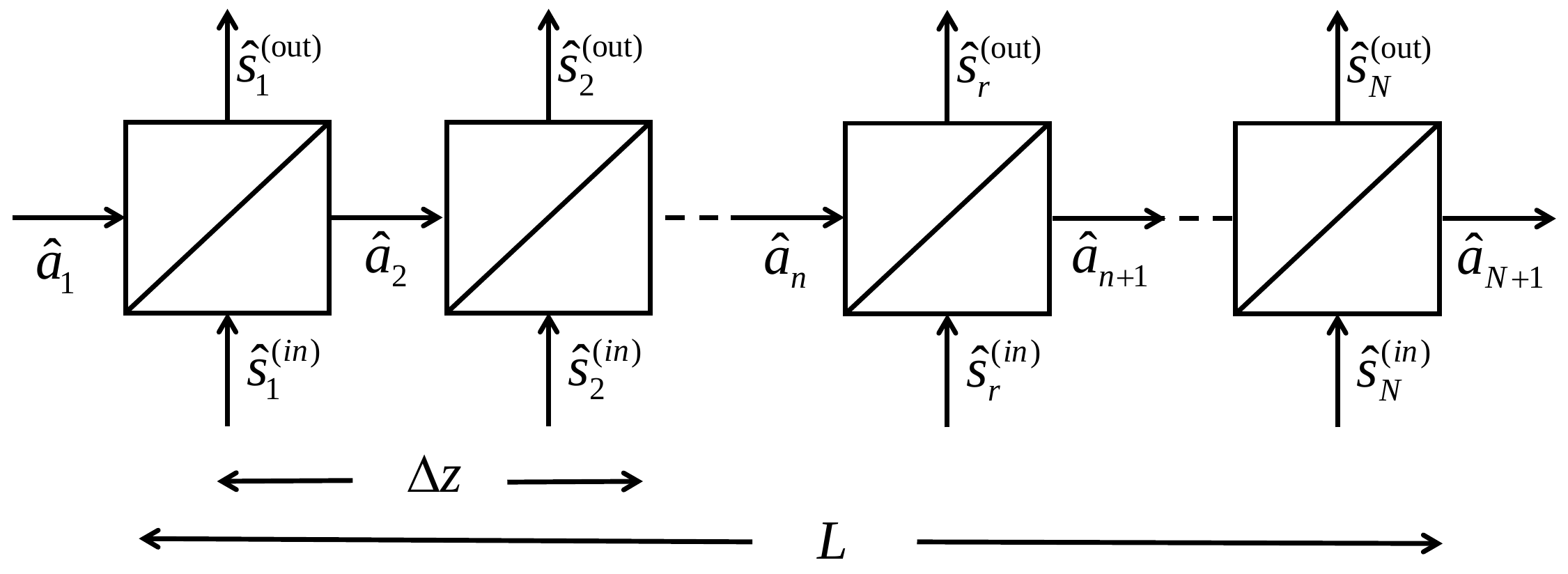} %&
%\includegraphics[width=3.0in,height=2.5in]{all_thru_RR_a_c}
%\end{tabular}
\caption{Loudon's traveling-wave attenuation by an infinite set of discrete beam splitters.
}\label{fig:loudon_bs_loss}
\end{figure}

To model loss in a quantized traveling wave field $\hat{a}$, Loudon considers successive propagation through an infinite series of fictitious beam splitters as illustrated in \Fig{fig:loudon_bs_loss}. For the $r$th beam splitter, $\hat{s}^{(in)}_r$ represents noise that is scattered \textit{into} the beam by scattering centers,
%=======================================================
%==================================
while $\hat{s}^{(out)}_r$ represents light that is scattered out of the beam. Each beam splitter (i.e. scattering center) is modeled by a frequency dependent transmission and reflection coefficient $T(\omega),\, R(\omega)$, respectively such that,
\bsub
\bea{eqn:BSr}
\hat{a}_{r+1}(\omega)     &=&  T(\omega)\,\hat{a}_{r}(\omega) + R(\omega)\,\hat{s}^{(in)}_r(\omega), \label{eqn:BS:ar}\\
\hat{s}^{(out)}_r(\omega) &=&  R(\omega)\,\hat{a}_{r}(\omega) + T(\omega)\,\hat{s}^{(in)}_r(\omega).
\eea
\esub
Here we assume that the pairs of input and output operators satisfy the usual boson commutation relations,
\be{comm:rels:discrete}
[\hat{a}_{r}(\omega), \hat{a}^\dagger_{r}(\omega')] =
[\hat{a}_{r+1}(\omega), \hat{a}^\dagger_{r+1}(\omega')] =
[\hat{s}^{(out)}_r(\omega), \hat{s}^{\dagger(out)}_r(\omega')] = \delta(\omega-\omega').
\ee
and that operators for the different scattering sites are independent and obey,
\be{comm:rels:sin:discrete}
[\hat{s}^{(in)}_{r}(\omega), \hat{s}^{\dagger(in)}_{r'}(\omega')] = \delta_{r,r'}\,\delta(\omega-\omega').
\ee

Successive iteration of  \Eq{eqn:BS:ar} yields,
\be{eqn:ar:discrete}
\hat{a}_{N+1}(\omega) =  T^N(\omega)\,\hat{a}_{1}(\omega) + R(\omega)\,\sum_{r=1}^{N} T^{N-r}(\omega)\,\hat{s}^{(in)}_{r}(\omega).
\ee
We now take the continuum limit $N\to\infty$, $\Delta z = L/N\to 0$ and $|R(\omega)|^2\to 0$ and define the attenuation constant
$\Gamma(\omega) = |R(\omega)|^2 / \Delta z$.
Using $|T(\omega)|^2 + |R(\omega)|^2=1$ we have,
\be{eqn:TN}
|T(\omega)|^{2N} = (1-|R(\omega)|^2)^N = (1-\Gamma(\omega)L/N)^N \to e^{-\Gamma(\omega) L},
\ee
for which we take,
\be{eqn:T}
T(\omega) =  e^{i\,n(\omega) (\omega/c)-\frac{1}{2} \Gamma(\omega)\,\Delta z} \equiv  e^{i\xi(\omega) \Delta z}, \quad
\xi(\omega) \equiv \beta(\omega) + i \Gamma(\omega)/2, \quad
\beta(\omega) \equiv n(\omega) (\omega/c).
\ee
In \Eq{eqn:T} we have chosen the phase of $T(\omega)$ to incorporate the free propagation constant $\beta(\omega) \equiv n(\omega) (\omega/c)$ through a
medium of index of refraction $n(\omega)$, and defined the complex propagation constant as $\xi(\omega) \equiv \beta(\omega) + i \Gamma(\omega)/2$.
We use $(N-r)\Delta z = L-z$ and convert from discrete to continuous modes through the identification,
\be{discrete:to:cont}
\hat{s}^{(in)}_{r}(\omega) \to (\Delta z)^{1/2}\hat{s}(z,\omega), \quad
\delta_{r,r'}\to \Delta z\,\delta(z-z'),
\ee
with commutation relations,
\be{comm:rels:s:cont}
[\hat{s}(z,\omega),\hat{s}^\dagger(z',\omega')] = \delta(z-z')\,\delta(\omega-\omega').
\ee
The continuous noise operators $\hat{s}(z,\omega)$ are assigned the expectation values,
\bsub
\bea{s:expect:values}
\langle \hat{s}(z,\omega) \rangle &=& \langle \hat{s}^\dagger(z,\omega) \rangle = 0,\\
\langle \hat{s}^\dagger(z,\omega) \, \hat{s}(z',\omega') \rangle &=& F_{\mathcal{N}}(\omega)\,\delta(z-z')\,\delta(\omega-\omega'),
\eea
\esub
where $F_{\mathcal{N}}(\omega)$ is the position-independent mean flux of noise photons per unit angular frequency.
Using $\sum_{r=1}^N\to (\Delta z)^{-1}\int_0^L dz$ we arrive at Loudon's expression for an attenuated traveling beam,
\be{aL}
\hat{a}_L(\omega) = e^{i\xi(\omega)L} \, \hat{a}_0(\omega) + i \sqrt{\Gamma(\omega)}\,\int_{0}^{L} dz \,e^{i\xi(\omega)(L-z)}\,\hat{s}(z,\omega),
\ee
where for convenience we have introduced the shorthand notation for the input field at $z=0\,$,  $\hat{a}_0(\omega)=\hat{a}(z,\omega)|_{z=0}$ and
the output field at $z=L\,$, $\hat{a}_L(\omega)=\hat{a}(L,\omega)$. Note that since $\hat{s}(z,\omega)$ are input noise operators, and $\hat{a}_0(\omega)$ is the input field before any interactions with the scattering centers, these operators commute,
\be{comm:rels:a:s}
[\hat{a}_0(\omega),\hat{s}(z',\omega')]= [\hat{a}_0(\omega),\hat{s}^\dagger(z',\omega')]= 0.
\ee
Thus, if we explicitly form the commutation relation $[\hat{a}_L(\omega), \hat{a}^\dagger_L(\omega')]$
we obtain two terms,
\bea{comm:rels:aL}
[\hat{a}_L(\omega), \hat{a}_L^\dagger(\omega')] &=& e^{i[\xi(\omega)-\xi^*(\omega')]L} \, [\hat{a}_0(\omega),\hat{a}_0(\omega')] \no
&+&  \sqrt{\Gamma(\omega)\Gamma(\omega')}\,\int_{0}^{L} dz \int_{0}^{L} dz'\,e^{i[\xi(\omega)(L-z)-\xi^*(\omega')(L-z')]}\,[\hat{s}(z,\omega), \hat{s}^\dagger(z',\omega')],\no
&=& \delta(\omega-\omega')
\big(\,
e^{-\Gamma(\omega) L} + \Gamma(\omega)\,\int_{0}^{L} dz \,e^{-\Gamma(\omega) z}
\,\big), \no
&=& \delta(\omega-\omega'),
\eea
where in the second equality we have used $i[\xi(\omega)-\xi^*(\omega')] = -\Gamma(\omega)$,
the commutation relations for $a_0(\omega)$ in \Eq{comm:rels:discrete}, and $s(z,\omega)$ in \Eq{comm:rels:s:cont} and that the integral in the second to last
line yields $(1-e^{-\Gamma(\omega) L})/\Gamma$. Thus, the expression for the attenuated traveling wave $\hat{a}_L(\omega)$ in \Eq{aL} explicitly preserves the output field commutation relations.
We can rewrite \Eq{aL} in a Langevin form as,
\bsub
\bea{aL:f}
\hat{a}_L(\omega) &=&   e^{i\xi(\omega)L} \, \hat{a}_0(\omega) + i \sqrt{1-e^{-\Gamma(\omega) L}}\, \hat{f}(\omega), \label{defn:aL:f}\\
\hat{f}(\omega) &\equiv& \frac{1}{\sqrt{1-e^{-\Gamma(\omega) L}}} \,\int_{0}^{L} dz \, e^{i\xi(\omega)(L-z)}\,\hat{s}(z,\omega), \label{defn:f:s}
\eea
\esub
where the Langevin noise operators $\hat{f}(\omega)$ satisfy the delta correlated commutation relations,
\be{}
[\hat{f}(\omega), \hat{f}^\dagger(\omega')] = \delta(\omega-\omega').
\ee

Note, that in the absence of loss $\Gamma=0$ \Eq{aL:f} reduces to the
un-attenuated free propagating field expression $\hat{a}_L(\omega) =   e^{i\beta(\omega)L} \, \hat{a}_0(\omega)$, which is unitary
since $|e^{i\beta(\omega)L}|=1$. One could deduce  \Eq{defn:aL:f} by phenomenologically introducing loss as
$\hat{a}_L(\omega) \sim   e^{[i\beta(\omega) -\Gamma(\omega)] L} \, \hat{a}_0(\omega)$, assuming $\hat{a}_L(\omega)$ takes the form
of $\hat{a}_L(\omega) = \mathcal{A}\,\hat{a}_0(\omega) + \mathcal{B}\,\hat{f}(\omega)$, with
$[\hat{f}(\omega), \hat{f}^\dagger(\omega')] = \delta(\omega-\omega')$,
and \textit{requiring} by quantum mechanics that
$[\hat{a}_L(\omega), \hat{a}_L^\dagger(\omega')] = \delta(\omega-\omega')$, which implies that $|\mathcal{B}| = \sqrt{1-|\mathcal{A}|^2}$ with freedom
to choose the phase of $\mathcal{B}$. This deduction is the essence of the Langevin approach, where the inclusion of loss requires the introduction of
additional noise operators $\hat{f}(\omega)$ to ensure that the quantum mechanical commutation relations
are preserved. What is not obtained from this procedure is the functional from
of $\hat{f}(\omega)$ as given by \Eq{defn:f:s}. The above derivation of $\hat{a}_L(\omega)$ by Loudon preserves the commutation relations
$[\hat{a}_L(\omega), \hat{a}^\dagger_L(\omega')] = \delta(\omega-\omega')$ by explicit construction.
%rather than by ansatz.

In the derivation of \Eq{aL} and subsequent commutation relation \Eq{comm:rels:aL}
a single loss $\Gamma$ was assumed throughout the whole length $L$ of the ring resonator.
This was not an essential assumption. If the ring resonator  had loss $\Gamma_1$ over length $L_1$ and loss $\Gamma_2$ over the remaining length $L_2 = L - L_1$ one can easily derive
\bea{aLLprime}
\hat{a}_L(\omega) &=& e^{i\xi_2(\omega)L_2} \, e^{i\xi_1(\omega)L_1} \, \hat{a}_0(\omega) \
+ e^{i\xi_2(\omega)L_2} \, i \sqrt{\Gamma_1(\omega)}\,\int_{0}^{L_1} dz \,e^{i\xi_1(\omega)(L_1-z)}\,\hat{s}(z,\omega)\no
&+& i \sqrt{\Gamma_2(\omega)}\,\int_{L_1}^{L} dz \,e^{i\xi_2(\omega)(L-z)}\,\hat{s}(z,\omega).
\eea
The commutation relation then yields a sum of terms given by (compare to \Eq{comm:rels:aL})
\bea{comm:rels:aLLprime}
[\hat{a}_L(\omega), \hat{a}_L^\dagger(\omega')]
&=& \delta(\omega-\omega')
\left(\,
e^{-\Gamma_2(\omega) L_2}\, e^{-\Gamma_1(\omega) L_1} \right. \no
&+& \left. e^{-\Gamma_2(\omega) L_2}\,\Gamma_1(\omega)\,\int_{0}^{L_1} dz \,e^{-\Gamma_1(\omega)(L_1-z)}
+ \Gamma_2(\omega)\,\int_{L_1}^{L} dz \,e^{-\Gamma_2(\omega) (L-z)}
\,\right), \no
&=& \delta(\omega-\omega').
\eea
%
%%
%\bea{comm:rels:aLLprime_v2}
%[\hat{a}_L(\omega), \hat{a}_L^\dagger(\omega')]
%&=& \delta(\omega-\omega')
%\big(\,
%e^{-\Gamma_2(\omega) L_2}\, e^{-\Gamma_1(\omega) L_1} \no
%%
%&+& e^{-\Gamma_2(\omega) L_2}\,\Gamma_1(\omega)\,\int_{0}^{L_1} dz \,e^{-\Gamma_1(\omega) z}
%+ \Gamma_2(\omega)\,\int_{L_1}^{L} dz \,e^{-\Gamma_2(\omega) (L-z)}
%\,\big), \no
%%
%&=& \delta(\omega-\omega').
%\eea
%
This result can be straightforwardly generalized to an arbitrary number of sections of the ring resonator of length $L_i$ with corresponding losses $\Gamma_i$ such that $\sum_i L_i = L$.
%============================
\section{Derivation of single bus commutation relation \Eq{c:comm:pma}}\label{app:pma:comm:deriv}
%============================
In \Eq{pma:deriv:7} we derived an expression for the output field $\hc$ in terms of
the input field $\ha$ and ring resonator noise operators $\hs(z,\om)$,
\be{pma:deriv:7:app}
\hc(\om) =
\left(
\frac{\t-\alpha\,e^{i\theta}}{1-\ts\,\alpha\,e^{i\theta}}
\right)\,\ha(\om)
-i |\k|^2\,\sqrt{\Gamma}\sum_{n=0}^\infty (\ts)^n\,\Int{(n+1)}
\ee
where  we have use the definition,
$a_{n+1} = e^{i\xi (n+1) L}\ha_0 + \Int{(n+1)}$ with
$\xi(\om) = \beta(\om) + i \Gamma(\om)/2$
such that $e^{i\xi L}  \equiv \alpha\,e^{i\theta}$ with  $\alpha=e^{-\frac{1}{2}\,\Gamma L}$
and $\theta = \beta L$.
In this appendix we wish to show explicitly that
output field commutation relation \Eq{c:comm:pma} yields,
\be{c:comm:pma:app}
[\hc(\om),\hc^\dagger(\om')]=\delta(\om-\om'),
\ee
where the input field and noise operators satisfy,
\be{comm:rels:a:s:app}
[\hat{a}(\omega),\hat{a}^\dagger(\omega')] = \delta(\om-\om'), \quad
[\hat{s}(z,\omega),\hat{s}^\dagger(z',\omega')]= \delta(z-z')\,\delta(\om-\om'),
\ee
and
\be{comm:rels:aa:ss:app}
[\hat{a}(\omega),\hat{s}(z',\omega')]
 = [\hat{a}(\omega),\hat{s}^\dagger(z',\omega')]= 0.
\ee

Let us define,
\be{A:atoc}
\mathcal{A}_{a\to c}  =
\left(
\frac{\t-\alpha\,e^{i\theta}}{1-\ts\,\alpha\,e^{i\theta}}
\right)
\equiv
e^{i\theta_\t}\,
\left(
\frac{|\t|-\alpha\,e^{i\theta'}}{1-|\t|\,\alpha\,e^{i\theta'}}
\right),
\quad
\t=|\t| e^{i\theta_\t}, \;\; \theta'\equiv\theta-\theta_\t
\ee
where we have defined $\t=|\t| e^{i\theta_\t}$ and the total phase angle $\theta'\equiv\theta-\theta_\t$,
so that we can write \Eq{pma:deriv:7:app} as,
\be{}
\hc(\om) = \mA_{a\to c}\,\ha(\om) -i\,\hF(\om).
\ee
The goal is to then show that,
\be{comm:F:defn}
[\hF(\om),\hF^\dagger(\om')] = (1 - |\mA_{a\to c}|^2)\,\delta(\om-\om').
\ee

Forming the commutator \Eq{c:comm:pma:app} we obtain,
\be{comm:Inm}
[\hat{c}(\omega),\hat{c}^\dagger(\omega')] = \delta(\om-\om')\,
\left(
|\mA_{a\to c}|^2 + \sum_{n=0}^{\infty}\,\sum_{m=0}^{\infty}\,I_{n,m}
\right),
\ee
where we have defined,
\be{Inm:defn}
I_{n,m} = \Gamma\,|\k|^4
(\ts)^n\,\t^m\,
 \int_{0}^{(n+1) L}dz\,\int_{0}^{(m+1) L} dz'\,
e^{i\xi(\omega)[(n+1) L - z]}\,e^{-i\xi^*(\omega')[(m+1) L - z']}\, \delta(z-z'),
\ee
where the spatial delta function in \Eq{Inm:defn} arises from using the commutators for the noise operators $\hat{s}(z,\omega)$ in \Eq{comm:rels:a:s:app}.
The last term in \Eq{comm:Inm} can be written as,
\be{sum:Inm}
\sum_{n=0}^{\infty}\,\sum_{m=0}^{\infty}\,I_{n,m}
=
\sum_{n=0}^{\infty}\,I_{n,n} +
2\,\sum_{n=0}^{\infty}\,\sum_{m=0}^{n-1}\,\textrm{Re}(I_{n,m}),
\ee
where we have used $I_{m,n}=I^*_{n,m}$.
The diagonal sum in \Eq{sum:Inm} is straightforwardly computed as,
\bea{sum:Inn}
\sum_{n=0}^{\infty}\,I_{n,n} &=&
\Gamma\,|\k|^4\,
\sum_{n=0}^{\infty}\,|\t|^{2 n}\,\int_{0}^{(n+1) L} dz \,e^{-\Gamma[(n+1)L-z]}, \no
&=&
|\k|^4\sum_{n=0}^{\infty}\,|\t|^{2 n}\,(1 - (\alpha^2)^{n+1}),\no
&=&
\frac{|\k|^2(1-\alpha^2)}{1-|\t|^2\,\alpha^2},
\eea
where we have used
$i(\xi - \xi^*) = -\Gamma$
and $\alpha^2 = e^{-\Gamma L}$.
For the off-diagonal sum in \Eq{sum:Inm} we use the fact that for $n>m$ and for some arbitrary function $f(z,z')$,
\be{int:delta}
\int_{0}^{(n+1) L} dz \int_{0}^{(m+1) L} dz' f(z,z') \, \delta(z-z')
= \int_{0}^{(m+1) L} dz' f(z',z'),
\ee
since the intergration over the longer interval $(n+1) L$ ensures the contribution of
the delta function on the shorter interval $(m+1) L$. We then obtain,
\be{sum:Re:Inm:1}
2\,\sum_{n=0}^{\infty}\,\sum_{m=0}^{n-1}\,\textrm{Re}(I_{n,m})
=
2\,|\k|^4\,\alpha^2\,\sum_{n=0}^{\infty}\,(\ts\alpha e^{i\theta})^n\,
\sum_{m=0}^{n-1}\, (\t\alpha e^{-i\theta})^m\,
\left(\frac{1}{(\alpha^2)^{m+1}}-1\right)
\ee
where $e^{i\xi L}\equiv \alpha\, e^{i\theta}$.
The above finite and infinite geometric sums can be computed using
$\sum_{m=0}^{n-1}\,x^m = (1-x^n)/(1-x)$ and
$\sum_{n=0}^{\infty}\,x^n = 1/(1-x)$.
%
%Without loss of generality we take $\t$ real,
%We now define $\t=|\t| e^{i\theta_\t}$ and the total phase angle $\theta'\equiv\theta-\theta_\t$.
After some lengthy but straightforward algebra one obtains,
\be{sum:Re:Inm:2}
2\,\sum_{n=0}^{\infty}\,\sum_{m=0}^{n-1}\,\textrm{Re}(I_{n,m})
=
2\,\frac{|\k|^2\,(1-\alpha^2)}{1-|\t|^2\,\alpha^2}\,
\frac{(|\t|\alpha\cos\theta' - |\t|^2\alpha^2)}{|1-|\t|\,\alpha\,e^{i\theta'}|^2}.
\ee
Adding \Eq{sum:Inn} to \Eq{sum:Re:Inm:2} yields,
\bsub
\bea{sum:Inm:final}
\sum_{n=0}^{\infty}\,\sum_{m=0}^{\infty}\,I_{n,m} &=&
\frac{|\k|^2 (1-\alpha^2)}{|1-|\t|\,\alpha\,e^{i\theta'}|^2}, \\
&\equiv& 1-|\mA_{a\to c}|^2,
\eea
\esub
where the last line follows from the use of the expression for $\mA_{a\to c}$
in \Eq{A:atoc} and $|\t|^2 + |\k|^2=1$.

Finally, the commutation relation \Eq{c:comm:pma:app} can be extended (though the algebra would be somewhat tedious) to the case of a ring resonator with an arbitrary number of sections  of length $L_i$ with corresponding losses $\Gamma_i$ such that $\sum_i L_i = L$ by using the results and generalizations of \Eq{aLLprime} and \Eq{comm:rels:aLLprime} at the end of the previous appendix.

%============================
%============================
%\newpage
\begin{acknowledgments}
PMA, AMS and CCT would like to acknowledge support of this work from OSD ARAP QSEP program.
EEH would like to acknowledge support for this work was provided by the Air Force Research
Laboratory (AFRL) Visiting Faculty Research Program
(VFRP) SUNY-IT Grant No. FA8750-13-2-0115.
%Some of the authors were also supported by the Air Force Office
%of Scientific Research under Grant No. FA9550-10-1-0217.
The authors also wish to thank M. Raymer for pointing out their previous related work.
Any opinions, findings and conclusions or recommendations
expressed in this material are those of the author(s) and do not
necessarily reflect the views of AFRL.
\end{acknowledgments}
%============================

%--------------------------------------------
%\newpage
\bibliography{rr_losses_refs}

\begin{thebibliography}{56}
\bibitem [{eps()}]{epsilon:note}%
  \BibitemOpen
  \href@noop {} {}\bibinfo {note} {The choice of $\epsilon=0.001$ for $0\le
  P^{(\alpha)}_{c,d}\le\epsilon$ was chosen to be ten times smaller than
  current high accuracy experimental detection realizations.}\BibitemShut
  {Stop}%
\end{thebibliography}
%-------------------------------------------
%\begin{thebibliography}{99}
%%
%\bibitem{Wei_Norman:1963} J. Wie and E. Norman "Lie algebraic soluition of linear differential equations," J. Math. Phys. {\textbf{4}}, 575  (1963).
%%
%\bibitem{Agarwal:2013} G.S. Agarwal, {\textit{Quantum Optics}}, Cambridge Univ. Press (2013).
%%
%\end{thebibliography}
%--------------------
\end{document}